\documentclass[iop]{emulateapj}

\usepackage{amsmath,amssymb}

\newcommand{\up}[1]{$^{#1}$}
\newcommand{\K}{~K}
\newcommand{\km}{~km}
\newcommand{\kms}{~km~s$^{-1}$}
\newcommand{\pcgs}{~dyn~cm$^{-2}$}
\newcommand{\Vturb}{$V_{\rm turb}$}
\newcommand{\Vradial}{$V_{\rm radial}$}
\newcommand{\intensity}{erg s\up{-1} cm\up{-2} sr\up{-1} \AA\up{-1}}
\newcommand{\Ek}{$E_{\rm k}$}
\newcommand{\Eh}{$E_{\rm h}$}
\newcommand{\Rkh}{$R_{\rm kh}$}
\newcommand{\Wk}{$W_{\rm k}$}
\newcommand{\VLOSk}{$V_{\rm LOS, k}$}

\shorttitle{Observations and Modelings of Limb Spicules}
\shortauthors{Tei et al.}

\slugcomment{Accepted 2019 November 26th}

\begin{document}

\title{IRIS Mg II Observations and Non-LTE Modeling
	\\ of Off-limb Spicules in a Solar Polar Coronal Hole}

\author{
Akiko Tei\altaffilmark{1,2}, 
Stanislav Gun\' ar\altaffilmark{3}, 
Petr Heinzel\altaffilmark{3}, 
Takenori J. Okamoto\altaffilmark{4}, \\
Ji\v r\' i \v St\v ep\' an\altaffilmark{3}, 
Sonja Jej\v ci\v c\altaffilmark{5,6}, and 
Kazunari Shibata\altaffilmark{1}}
\email{teiakiko@kwasan.kyoto-u.ac.jp}
\altaffiltext{1}{Astronomical Observatory, Graduate school of Science, Kyoto University, Kyoto 607-8471, Japan}
\altaffiltext{2}{Research Fellow of Japan Society for the Promotion of Science, Tokyo 102-0083, Japan}
\altaffiltext{3}{Astronomical Institute, Academy of Sciences of the Czech Republic, 25165 Ond\v rejov, Czech Republic}
\altaffiltext{4}{National Astronomical Observatory of Japan, Mitaka, Tokyo 181-8588, Japan}
\altaffiltext{5}{Faculty of Education, University of Ljubljana, 1000 Ljubljana, Slovenia}
\altaffiltext{6}{Faculty of Mathematics and Physics, University of Ljubljana, 1000 Ljubljana, Slovenia}

\begin{abstract}
We investigated the off-limb spicules observed in the Mg II h and k lines by IRIS in a solar polar coronal hole.
We analyzed the large dataset of obtained spectra to extract quantitative information about the line intensities, shifts, and widths.
The observed Mg II line profiles are broad and double-peaked at lower altitudes, broad but flat-topped at middle altitudes, and narrow and single-peaked with the largest Doppler shifts at higher altitudes.
We use 1D non-LTE vertical slab models (i.e. models which consider departures from Local Thermodynamic Equilibrium) in single-slab and multi-slab configurations to interpret the observations and to investigate how a superposition of spicules along the line of sight (LOS) affects the synthetic Mg II line profiles.
The used multi-slab models are either static, i.e. without any LOS velocities, or assume randomly assigned LOS velocities of individual slabs, representing the spicule dynamics.
We conducted such single-slab and multi-slab modeling for a broad set of model input parameters and showed the dependence of the Mg II line profiles on these parameters.
We demonstrated that the observed line widths of the h and k line profiles are strongly affected by the presence of multiple spicules along the LOS.
We later showed that the profiles obtained at higher altitudes can be reproduced by single-slab models representing individual spicules.
We found that the multi-slab model with a random distribution of the LOS velocities ranging from -25 to 25\kms\ can well reproduce the width and the shape of Mg II profiles observed at middle altitudes. 
\end{abstract}

\section{INTRODUCTION} \label{s-intro}
Solar spicules remain an enigmatic part of the solar atmosphere, with some of their properties being still not well understood. 
This is due to the fact that the dimensions and temporal variations of spicules are near the resolution limits of the current observations. 
At the same time, the ubiquitous nature of spicules means that in observations many of them are often present along the line of sight (LOS). 
The high number of concurrently existing spicules and a distribution of their anchorage along supergranules' edges means that at low altitudes above the solar limb, despite the broad distribution of inclinations and plasma velocities of the spicules, the probability of observing multiple spicules along a LOS is very high.
On the other hand, due to the above-mentioned distribution of the inclinations and broad distributions of plasma velocities as well as the lengths of the spicules, one can often observe individual spicules at the top and above the top of spicules' forest.

\par
We present here only a short overview of the properties of spicules. 
These were comprehensively reviewed by \citet{bec68} and later by \citet{tsi12}.
Since then, additional information about spicules was published in works based on observations by Interface Region Imaging Spectrograph \citep[IRIS,][]{dep14}. 
The term ``spicules'' traditionally refers to thin, elongated, short-lived, dynamic structures observed above the limb, reaching altitudes of up to 10~Mm.
The dynamic nature of spicules can be characterized by the observed motions in the direction along the spicules and also in the transverse direction.
Motions along the spicules have typical velocities ranging from 15 to 40\kms, but sometimes reaching up to over 100\kms\ \citep[see e.g.][]{dep07, zha12, per12}.
These velocities are typically measured from high-cadence imaging observations, such as those obtained by the Solar Optical Telescope \citep[SOT,][]{tsu08} on-board the Hinode satellite \citep{kos07}.
These observational analyses are focused on individually well-resolved spicules reaching higher altitudes. 
As such, a selection effect favoring spicules with a certain set of properties may be at play \citep{per12}.
The measurements of the velocities are performed in the observed plane of the sky and the obtained values thus represent a projection of the actual velocities onto the observational plane. 
The observed velocities perpendicular to the main axis of spicules are between 5 and 30\kms\ \citep{dep07, per12}. 
For example, \cite{dep12} distinguish two types of transverse motions: swaying motions (15-20\kms) and torsional motions (25-30\kms) 
-- see also \citet{dep07}, who reported that the solar chromosphere is permeated by Alfv\' en waves with amplitudes of the order of 10-25\kms\ and periods of 100-500~s.
Due to the inclination of spicules with respect to the radial direction (\cite{per12} for example reported mean apparent inclination of about $\pm$25 deg from the vertical), both the motions along spicules and the transverse motions result in LOS velocities of individual observed spicules.
The observed unsigned LOS velocities range from less than 10\kms\ \citep[e.g.][]{pas09} to 30\kms\ \citep[e.g.][]{sko14}.

\par
Temperatures and electron densities of the spicule plasma are still not fully determined \citep[see][]{tsi12}. 
This is because such determination requires multi-line spectroscopic observations and employment of the non-LTE (i.e. departures from Local Thermodynamic Equilibrium) radiative transfer modeling.
These thermodynamic properties of spicules were previously analyzed by \cite{bec72}, whose results indicate that the temperature of spicule plasma rises with the altitude above the limb (from 9,000~K at 2~Mm to 16,000~K at 8~Mm), while the electron density decreases (from 1.6 $\times$\ 10$^{11}$ cm$^{-3}$ at 2~Mm to 4.3 $\times$\ 10$^{10}$ cm$^{-3}$ at 8~Mm). 
\cite{kra71} obtained electron densities ranging from 10$^{11}$
cm$^{-3}$ to 10$^{12}$ cm$^{-3}$ at 5~Mm to 9~Mm. 
\cite{ali73} derived electron densities between 6 $\times$\ 10$^{10}$ cm$^{-3}$ and 1.2 $\times$\ 10$^{11}$ cm$^{-3}$, and temperatures between 10,000 and 15,000~K at altitude of 5.4~Mm. 
Later, \cite{kra76} found electron densities between 1.1 $\times$\
10$^{11}$ cm$^{-3}$ at 6 Mm and 2 $\times$\ 10$^{10}$ cm$^{-3}$ at
10 Mm, with temperature ranging from 12,000 to 15,000~K. 
These authors also determined the micro-turbulent velocities with values
between 12 and 22\kms. 
In contrast to \cite{bec72}, \cite{mat88} found more complex temperature variations with the altitude, giving temperature of 9,000 K at 2.2 Mm, 5,000 K at 3.2 Mm and 8,200~K at 6~Mm. 
\par
More recently, \cite{ali18} used time-averaged Mg II h and k, C II and Si IV spectra obtained by IRIS and employed single-slab 1D non-LTE modeling to analyze properties of quiet Sun, polar region spicules. 
These authors derived temperatures ranging from 8,000~K at the low
altitudes to 20,000~K at the top of spicules. 
Electron densities range from 1.1 $\times$\ 10$^{11}$ cm$^{-3}$ to 4 $\times$\ 10$^{10}$ cm$^{-3}$ and the micro-turbulent velocities reach rather high values of 24\kms. 
Similar IRIS observations of polar quiet-Sun spicules were analyzed by \cite{per14}, together with imaging data from Hinode/SOT. 
These authors derived velocities of ascending and descending phases of evolution to be around 80\kms\ upwards and as much as 140\kms\ downwards.

\par
In the present paper, we analyze high-cadence spectroscopic observations of off-limb spicules in a polar coronal hole obtained in Mg II h and k lines by IRIS (see Sec.~\ref{s-obs}).
We focus here on detailed analysis of LOS velocities, line-profile widths, integrated intensities of both Mg II lines and their ratio (see Sec.~\ref{ss-data} and \ref{ss-analysis}). 
We study here, for the first time, the evolution of the spicules through more than an hour of observations using over 37,000 individual spectra. 
The IRIS observations used here were obtained with very short exposure times of $\sim$ 5.4~s, which allows us to freeze the evolution stages of highly dynamical spicules within the short exposure. 
We can thus obtain information about immediate properties of spicules that are not affected by averaging over temporal variations. 
We also investigate the dependence of the measured LOS velocities, line widths and integrated intensities on the altitude above the solar disk. 
Such analysis was also performed by \cite{ali18}, who used temporally averaged Mg II h and k IRIS observations of quiet-Sun polar-region spicules.
We also use data in a polar region within the coronal hole, which is important for understanding of mass and energy transport between the cool, lower solar atmosphere, the hot, higher solar atmosphere, and the solar wind.

\par
To interpret the results of observational analysis, we use the non-LTE models employing 1D vertical slab geometry (Sec.~\ref{s-mod}).
We investigate the difference between single-slab models (Sec.~\ref{ss-single}) and models with multiple slabs along the LOS. 
We employ two configurations of multi-slab models -- a static multi-slab model (Sec.~\ref{ss-multi}) and a multi-slab model with randomly assigned LOS velocities for each slab (Sec.~\ref{ss-random}).
Such multi-slab models are used here for the first time to simulate off-limb spicules. 
The 1D non-LTE models used here were developed by \cite{hei14} to simulate solar prominences. 
These models, which solve the radiative transfer in hydrogen and Mg II spectral lines, were used for interpretation of prominence fine-structure observations also by \cite{hei15}, \cite{jec18} or Ruan et al. (2019) [ApJ acceptted].
Similarly, configurations of multiple 1D or 2D non-LTE models adopting vertical geometry were used for modeling of prominences and their fine structures by, e.g., \cite{Mor78} or \cite{gun07}. 
A 2D multi-thread model with random LOS velocities of individual prominence fine structures was developed by \cite{gun08}. 
In Sect.~\ref{s-para}, we show the dependence of the synthetic Mg II h and k line profiles on different parameters of the models. 
In Sect.~\ref{s-discussion}, we discuss the links between the results obtained by the analysis of the observed and synthetic spectra and in Sect.~\ref{s-conclusion}, we offer our conclusions. 

\section{OBSERVATIONS} \label{s-obs}
\subsection{Data} \label{ss-data}
\begin{figure*}[ht]
\centerline{\includegraphics[width=\linewidth]{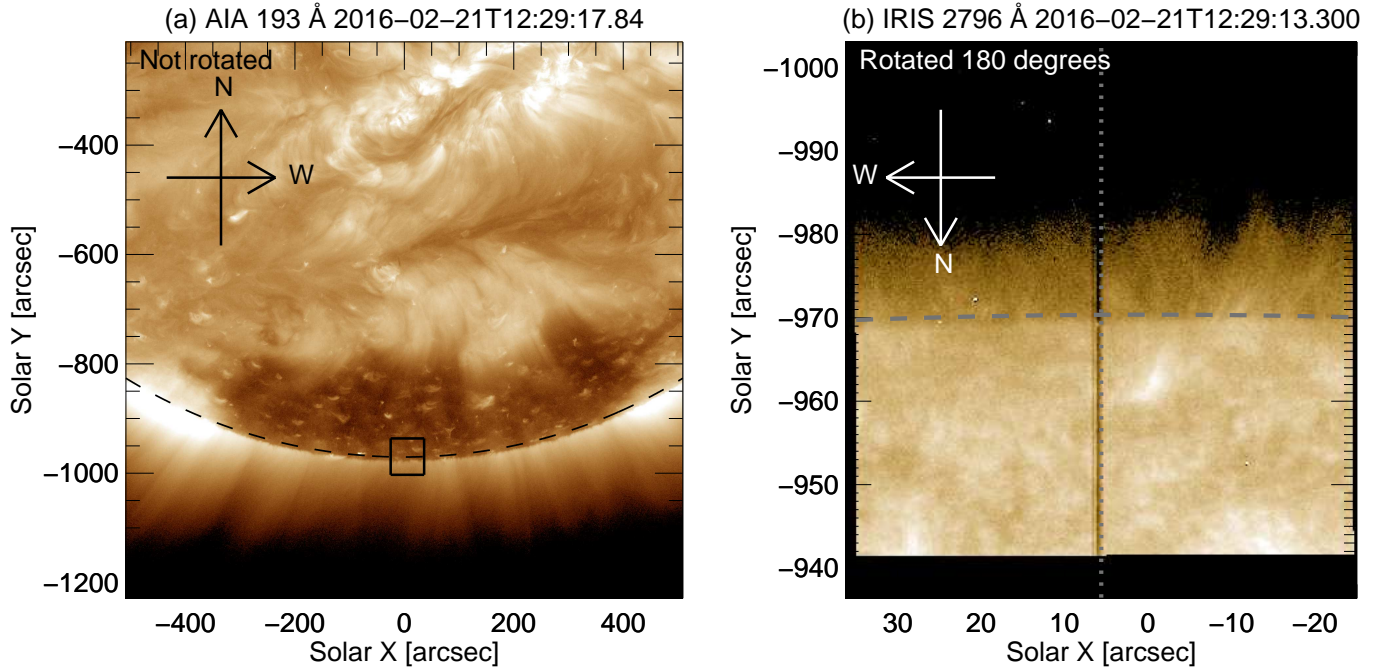}}
\caption{
Solar images recorded at the beginning of the IRIS observation. 
(a) The southern part of the Sun observed in AIA 193~\AA\ channel.
(b) IRIS slit-jaw image in 2796~\AA\ passband.
North is up and West is right in panel (a) while North is down and West is left in panel (b), where the image is rotated by 180 degrees.
Dashed curve in each panel represents the solar limb location from AIA index.
The small square in panel (a) shows the FOV of the IRIS slit-jaw image.
Dotted vertical line in panel (b) shows the IRIS slit location.
The observation time is shown for each image.
}
\label{fig-fov}
\end{figure*}

We used the data of a coronal hole in the southern polar region observed by IRIS (see Figure~\ref{fig-fov}).
The observation was conducted from 12:29 UT to 13:29 UT on 2016 February 21, in the medium sit-and-stare mode (OBS-ID: 3600257402). 
The slit was in the north-south direction.
The width of the IRIS slit is 0.33\arcsec. 
In the observation that we use in the present study, the length of the slit was 60\arcsec, the cadence of the spectral data was 5.4 s, the spatial pixel size along the slit was 0.17\arcsec, and the spectral pixel size was 25.6~m\AA.
In this work, we focused on the Mg II h and k spectra.
In Figure~\ref{fig-fov}(b), we show the IRIS slit-jaw image taken in the 2796~\AA\  passband with a field of view (FOV) of 60\arcsec\ $\times$\ 65\arcsec\ and the pixel size of 0.17\arcsec.
Note that the view in Figure~\ref{fig-fov}(b) is rotated by 180 degrees.

\par
To process the IRIS Mg II h and k data, we used the spatial and wavelength information in the header of the IRIS level-2 data and derived the rest wavelengths of the Mg II k and h as 2796.35~\AA\ and 2803.52~\AA\ from the reversal positions of the averaged spectra at the disk.
For our analysis, we rotated the spectral data by 180 degrees from the original north-south direction.
Therefore, in all spectra shown hereafter, the South is up and the North is down.

\par
In Figure~\ref{fig-fov}, we show the overview of the observed region.
From panel (a), it is clear that the southern polar region was a coronal hole at the time of the observations.
This panel shows a 193~\AA\ image obtained by the Atmospheric Imaging Assembly \citep[AIA;][]{lem12} on board the Solar Dynamics Observatory \citep[SDO;][]{pes12}.
The limb location from the AIA header information is indicated by a dashed curve in both panels of Figure~\ref{fig-fov}.
The FOV of the IRIS slit-jaw image in the 2796~\AA\ passband (Figure~\ref{fig-fov}(b)) is indicated by a small square in Figure~\ref{fig-fov}(a).

\begin{figure*}[t]
\centerline{\includegraphics[width=\linewidth]{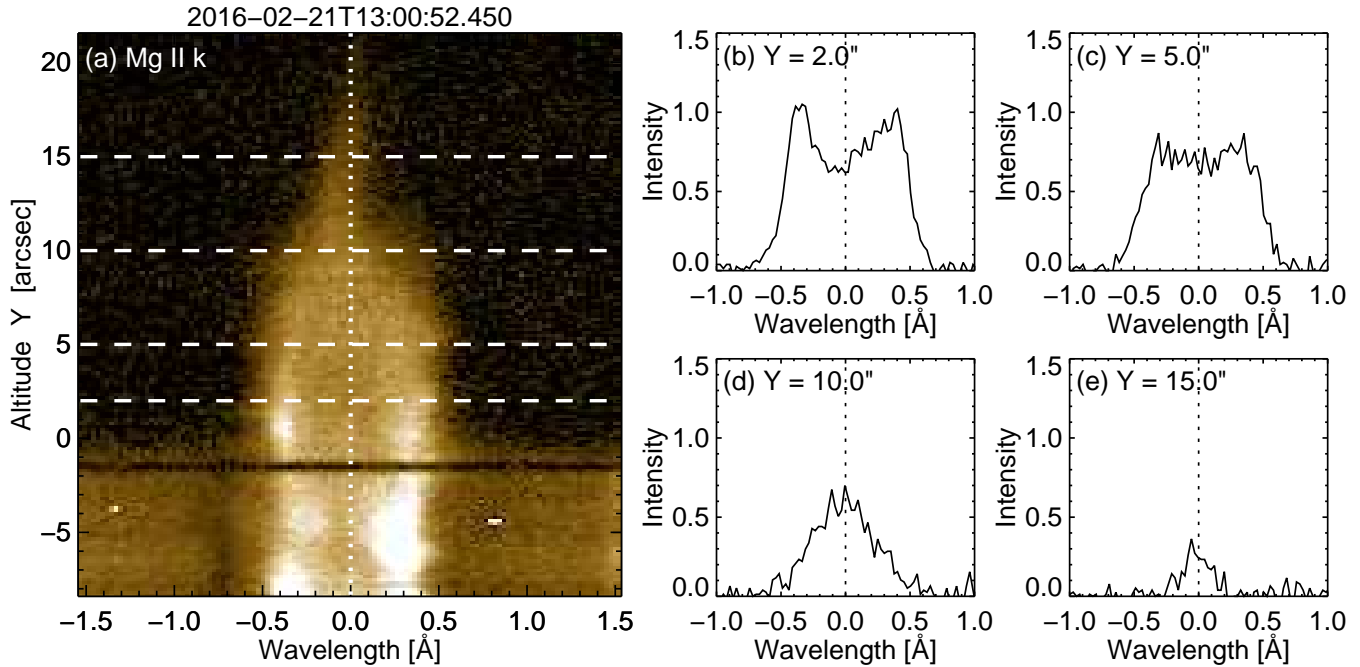}}
\caption{
The Mg II k spectrum observed by IRIS at 13:00:52. 
Four dashed lines in panel (a) show the altitudes represented in panels (b-e).
We show here the Mg II k line profiles at altitudes of $Y$ = 2.0\arcsec, 5.0\arcsec, 10.0\arcsec, and 15.0\arcsec.
Horizontal axis in each panel indicates the wavelength with respect to the line center of the Mg II k in the rest frame.
Dotted line in panels (b-e) shows the rest wavelength.
For time evolution, see the online movie (link here). 
}
\label{fig-irissp}
\end{figure*}

\begin{figure*}[!ht]
\centerline{\includegraphics[height=0.85\linewidth,angle=-90]{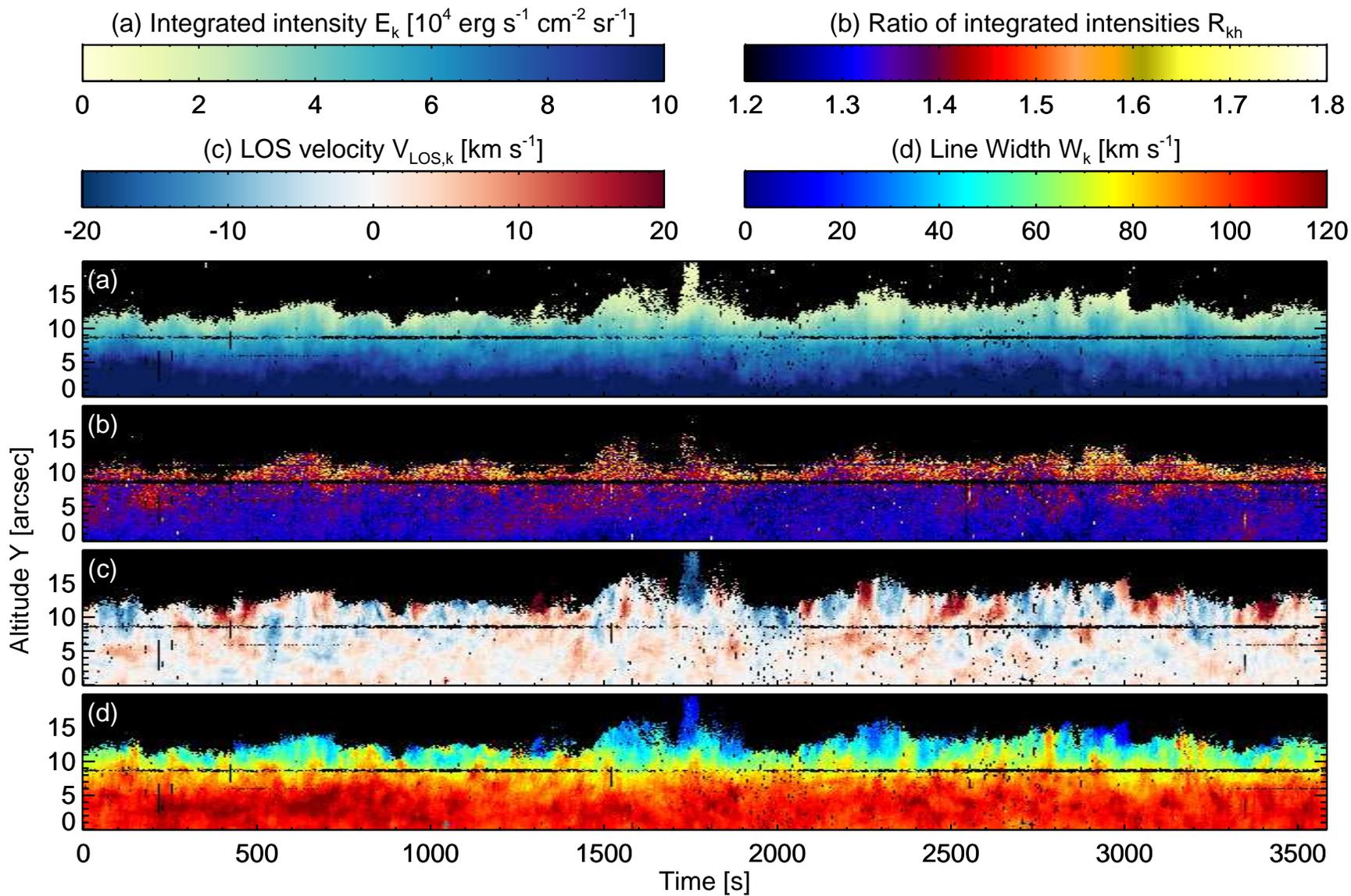}}
\caption{
Space-time plots of 
(a) integrated line intensities of the Mg II k line \Ek, 
(b) the ratios of the integrated line intensities in the Mg II k and h lines \Rkh (= \Ek / \Eh), 
(c) LOS velocities \VLOSk, and 
(d) line widths \Wk.
The color scale is shown for each.
The horizontal axis starts at the beginning of the observations (12:29:07~UT).
}
\label{fig-map}
\end{figure*}

\begin{figure*}[ht]
\centerline{\includegraphics[width=\linewidth]{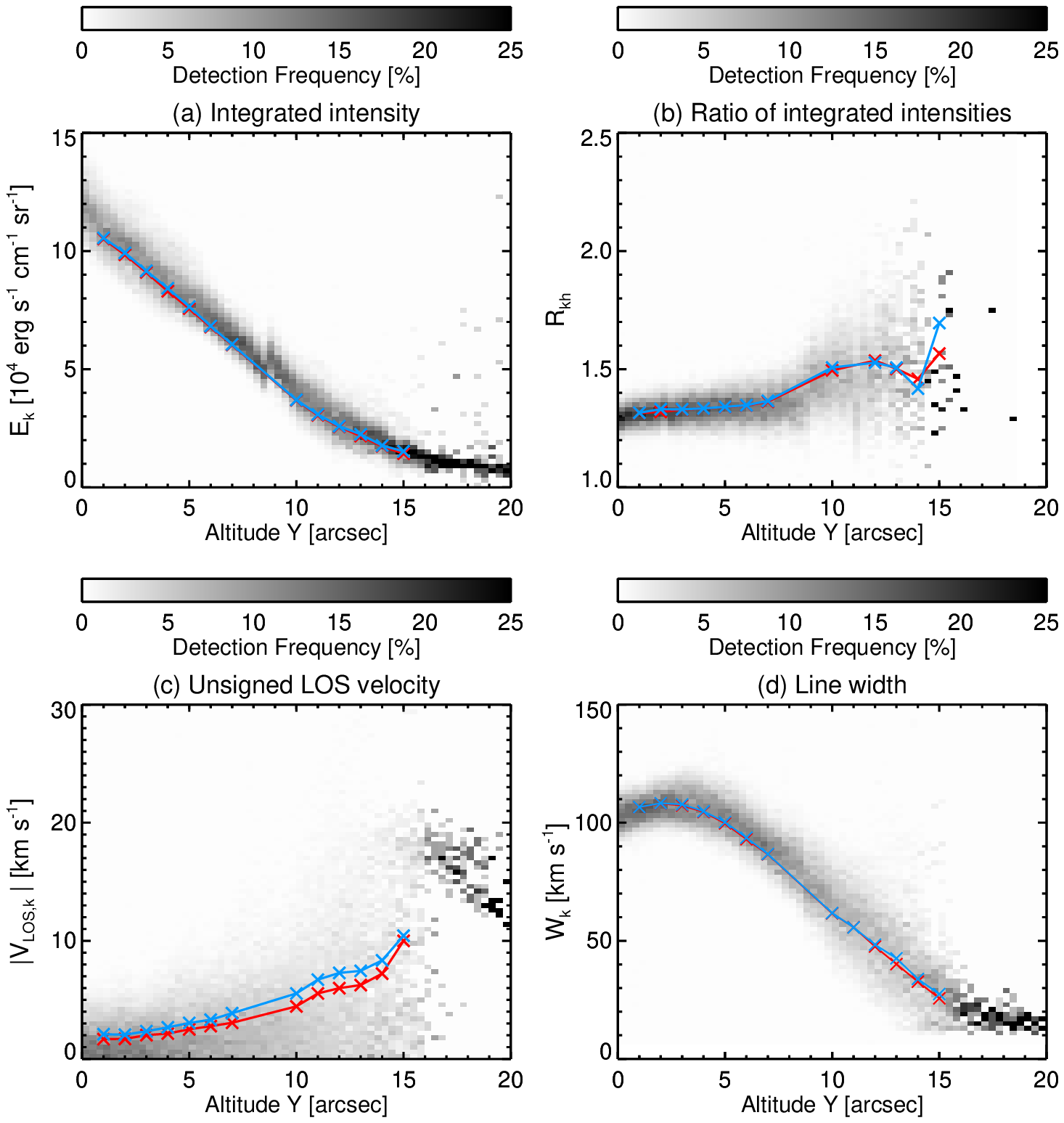}}
\caption{
Distribution of 
(a) integrated line intensities of the Mg II k line \Ek, 
(b) the ratio of the integrated line intensities in the Mg II k and h lines \Rkh (= \Ek /\Eh), 
(c) absolute values of the LOS velocities $|$\VLOSk$|$, and 
(d) line widths \Wk\ as a function of altitude $Y$.
The gray color shows the detection frequency at each altitude bin normalized at each altitude bin with a bin size of 0.33\arcsec.
The red line shows the median and blue line the mean of the plotted values at each altitude with an averaging bin size of 1\arcsec.
}
\label{fig-dist}
\end{figure*}

\subsection{Analysis} \label{ss-analysis}
In this section, we present the results of our analysis of the entire data set of the Mg II h and k observations described in the previous subsection.
This data set contains 660 exposures and, in an average exposure, spicules are covered by $\approx$70 pixels along the slit.
An example of the Mg II k spectrum is shown in Figure~\ref{fig-irissp}.
The studied Mg II line profiles show a complex behavior at different heights above the limb.
The line profiles at lower altitudes are broad and generally double-peaked (Figure~\ref{fig-irissp}(b)), profiles at middle altitudes are rather flat-topped (Figure~\ref{fig-irissp}(c)) and those at upper parts of spicules tend to be narrower and single-peaked (Figure~\ref{fig-irissp}(d,e)).
This complex behavior can be seen also in the online movie (link here) showing the entire observed time series.
In our analysis, we first converted the data count rate ([DN s\up{-1} pix\up{-1}]) to radiance ([\intensity]) using the effective area of NUV passband obtained from the IDL function ``iris\_get\_response.pro.''
Note that we adopted the spatial information in the header of the IRIS level-2 data and the SDO AIA 193\AA\ data for the spatial alignment between IRIS and AIA data.
We adopted the solar radius in the AIA header information to all data and defined the position of the solar radius on the IRIS slit as the altitude $Y$ = 0\arcsec.

\par
From the observed spectra in each exposure and at all pixels covering spicules, we extracted quantitative information about the integrated intensity, the shift of the line profile with respect to the rest wavelength, and the line width in the following way.
We measured the integrated intensity in both h and k lines (\Eh\ and \Ek).
In addition, we calculated the ratio of the integrated intensities in the h and k lines (\Rkh\ = \Ek/\Eh).
The measured integrated intensities in the Mg II k line and the line ratios are shown in Figure~\ref{fig-map}(a,b).
The black color means that the integrated intensity is too small to show.
The intensity \Ek\ is larger at the lower altitudes and smaller at the higher altitudes.
The upper part of tall spicules especially have small values of integrated intensity (e.g. the tall feature around time 1750 s).
The line ratio shows large value only for the highest part of spicules.
Note that data around $Y$ = 8.5\arcsec--9.0\arcsec\ are inadequate (probably due to some dust on the slit).
The distribution of these values as a function of altitude is shown in Figure~\ref{fig-dist}(a,b).
In this figure, we show the detection counts at each altitude bin normalized at each altitude bin.
Along the altitude we use a bin size of 0.33\arcsec, which is twice the size of the original binning size of observational data.
The red line shows the median and blue line the mean of the plotted values at each altitude with an averaging bin size of 1\arcsec.
The median and mean is not shown for $Y$ = 8\arcsec, 9\arcsec\ in Figure~\ref{fig-dist}(a,b) and for $Y$ = 11\arcsec\ in Figure~\ref{fig-dist}(b) due to the inadequate data.
The median and mean of the integrated intensities clearly decrease with altitude, from about 10$^5$ \intensity\ at $Y\sim$ 2\arcsec\ to about 1.5 $\times$ 10$^4$ \intensity\ at $Y\sim$ 15\arcsec.
The line ratio is nearly constant (around 1.3) below $Y$ = 7\arcsec\ but increase to values of up to 2 or more above $Y$ = 7\arcsec.


\par 
To measure the shift of the line profiles, we used the 50$\%$ bisector method.
In the bisector method, we defined $\lambda_{\rm blue}$ and $\lambda_{\rm red}$ as the two wavelengths at the 50$\%$ maximum intensity level, and then defined the line shift
as $\lambda_{\rm bis}=(\lambda_{\rm blue}+\lambda_{\rm red})/2$. 
The corresponding LOS velocity was defined as
$V_{\rm LOS}=c\ (\lambda_{\rm bis}-\lambda_{0})/\lambda_{0}$,
where $c$ is the speed of light and $\lambda_{0}$ is the rest wavelength.
The derived line shifts in the Mg II k line \VLOSk\ are shown in Figure~\ref{fig-map}(c).
The most important result of this analysis is that the derived LOS velocities are significantly smaller at lower altitudes of the observed spicules than at higher ones.
In addition, the derived LOS velocities at upper part of the observed spicules do not have distinct blue-red asymmetry.
To better compare the amplitude of the derived LOS velocities, we show in Figure~\ref{fig-dist}(c) the distribution of the unsigned values of the derived LOS velocities $|$\VLOSk$|$ as a function of altitude $Y$.
The displayed plots correspond to the LOS velocities shown in the time-distance plots in Figure~\ref{fig-map}(c).
The mean and median of the measured unsigned LOS velocities can be more than 10\kms\ at the higher altitudes while they are typically around 2\kms\ at the lower altitudes.
Figure~\ref{fig-dist}(c) also shows that the absolute values of LOS velocities can reach more than 20\kms\ at the higher altitudes, which is consistent with the velocity amplitude of transverse motions in spicules reported in \cite{dep07}.
This suggests that we observe a few or even individual spicules within the pixel in the higher altitudes.

\par 
In addition to the analysis of line shifts, we studied here the line widths using the 50$\%$ bisector method.
We defined the line width as 
$\Delta\lambda_{\rm bis}=(\lambda_{\rm red}-\lambda_{\rm blue})/2$ 
in the wavelength units and 
$W=c\times\Delta\lambda_{\rm bis}/\lambda_{0}$
in the velocity units.
The results of the analysis of the measured line widths in the Mg II k line \Wk\ are shown in time-distance plot in Figure~\ref{fig-map}(d).
Figure~\ref{fig-map}(d) shows clear stratification of the measured line widths with the altitude.
These results are clearly visible also in the distribution plots of the measured line widths as a function of the altitude, which we show in Figure~\ref{fig-dist}(d).
The line widths at lower altitudes are significantly greater (e.g. $\sim$100\kms; $\sim$0.9~\AA\ at $Y\sim$ 2\arcsec) than those measured at the upper part of spicules (e.g. $\sim$30\kms; $\sim$0.3~\AA\ at $Y\sim$ 15\arcsec).
It is also important to note that \Wk\ increased in the range $Y$ $<$ 3\arcsec\ and then decreased in the range $Y$ $>$ 3\arcsec\ toward higher altitudes, respectively.

\par
Note that in Figure~\ref{fig-map} we can clearly identify a very tall individual spicule in the upper part of the observed spectra at around 1750~s after the beginning of the observations (12:29:07~UT) and this feature is the main contributor to the dense distribution in the upper altitudes in Figure~\ref{fig-dist}.
Such a spicule would be easy to identify in high-resolution imaging observations.


\begin{figure*}[ht]
\centerline{\includegraphics[width=\linewidth]{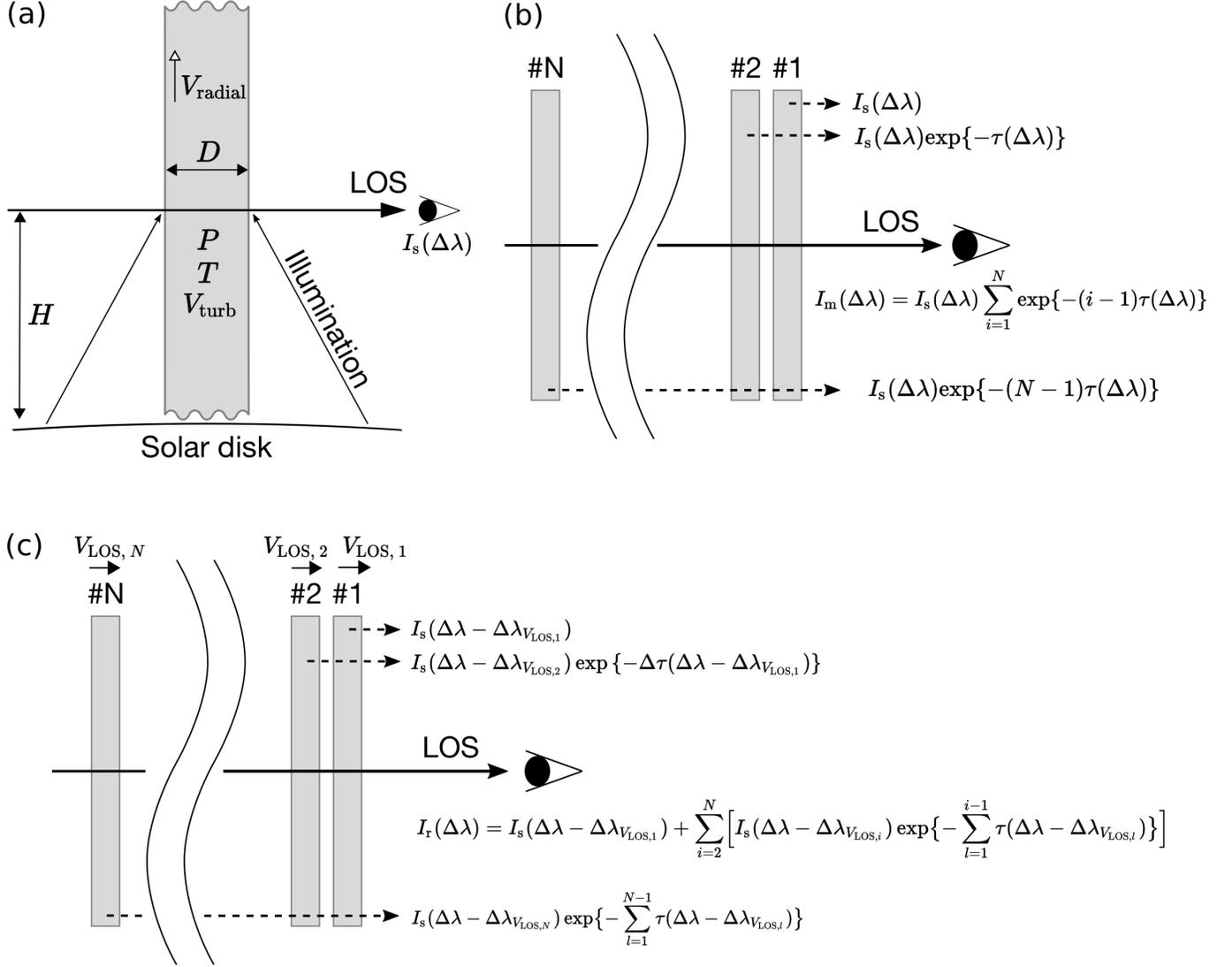}}
\caption{
Sketch of the (a) single-slab model, (b) multi-slab model without LOS velocities, and (c) multi-slab model with LOS velocities.
In panel (a), we show a single slab illuminated from the solar disk.
Here, $H$ is height above the solar surface, $P$ is pressure, $T$ is temperature, $D$ is geometrical thickness of the slab, \Vturb\ is micro-turbulent velocity, and \Vradial\ is radial velocity.
In panel (b), each dashed line indicates the intensity contributed by the slab to the computed intensity profile of the multi-slab model without LOS velocities.
In panel (c), each dashed line indicates the intensity contributed by the slab to the computed intensity profile of the multi-slab model with LOS velocities.
In each panel, the LOS direction of the calculation of the intensity line profile is indicated by the horizontal arrow pointing toward the observer.
}
\label{fig-model}
\end{figure*}

\clearpage 

\section{MODELING} \label{s-mod}
To interpret the results of our analysis of the observations mentioned above, we conducted the non-LTE radiative-transfer modeling in the following way.
First, we calculated the Mg II h and k line profiles from one-dimensional (1D) single-slab model (see Sec.~\ref{ss-single}).
Second, we experimented with additional slabs along the LOS to see how the profiles change when there are more than one slab (see Sec.~\ref{ss-multi}).
Third, we introduced a random LOS velocity for each slab of such a multi-slab model (see Sec.~\ref{ss-random}).
In the last subsection (Sec.~\ref{ss-convol}), we describe how the IRIS instrumental broadening affects the synthetic line profiles produced by the models.
We note that the instrumental broadening was applied to all synthetic line profiles shown in this work.


\subsection{Single-slab Model}\label{ss-single}
The 1D non-LTE radiative transfer code MALI \citep{hei14} was used here to synthesize the Mg II h and k line profiles from a single slab model.
The adopted 1D isothermal and isobaric slab stands vertically on the solar surface and has a finite geometrical thickness along a horizontal LOS (see Figure~\ref{fig-model}(a)).
The slab is illuminated at both sides from the solar disk.
Under the spicule conditions, this incident radiation plays a crucial role in determining the line source functions through the scattering.
For the Mg II incident radiation coming from the solar disk, we adopt the quiet-Sun observation from the balloon experiment RASOLBA \citep{sta95}, which is similar to that obtained by IRIS \citep[see Figure~4 in][]{liu15}.
We adopted results of this quiet-Sun observation as a model of the illumination from the disk, even though the spicule observations used here are from the polar coronal hole. 
We acknowledge that this assumption could cause some differences in the optical thicknesses, source functions and consequently in the specific intensities. 
However, such differences could be expected to be in a few tens of percent or less. This would not affect the overall results or conclusions of this paper.
We also adopt center-to-limb variations of the disk radiation.
The slab illumination varies with height above the surface due to the dilution factor.
In the present study, we used a fixed height above the surface, $H$~=~5000\km, which corresponds to the half of the typical maximum height of spicules.
However, the resulting synthetic profiles produced by the model do not depend significantly on the adopted height within the range of typical spicule altitudes.
The MALI code first solves the non-LTE problem for a 5-level plus continuum hydrogen atom and then for a 5-level plus continuum Mg II and Mg III ions.
Partial frequency redistribution (PRD) is considered for all hydrogen and magnesium resonance lines.
Input parameters in the model are the temperature $T$, gas pressure $P$, geometrical thickness $D$, micro-turbulent velocity \Vturb, and radial (vertical) velocity \Vradial.
\Vradial\ is important for calculations of the incident radiation due to the Doppler dimming/brightening effects \citep[see][]{hei14}.
By assuming specific values of these five input parameters, we obtain the ionization equilibrium of hydrogen and thus the electron densities, which are then used to solve the non-LTE problem for magnesium. Finally we get the synthetic intensities of Mg II lines together with their optical thicknesses.
We note that a 1D vertical slab approximates rather well the line source function of a cylindrical spicule. The difference in the line source function
between 1D slab and 1D axially-symmetric cylinder roughly amounts to 20 \%, as demonstrated by \cite{hea77}.
In the present work, we thus use the 1D slab approximation, which is computationally less demanding. 

\par
The optical thickness of a single-slab as a function of wavelength $\tau(\Delta\lambda)$ is calculated by considering thermal and non-thermal broadening and damping parameters of the natural, Stark, and Van der Waals broadenings.
The mean thermal velocity $V_{\rm th}$ that corresponds to the slab temperature $T$ and the (non-thermal) micro-turbulent velocity \Vturb\ of the slab were adopted to calculate the total Doppler width in frequency unit $\Delta\nu_D=(\nu_0/c)\sqrt(V_{\rm th}^2+V_{\rm turb}^2)$, where $\nu_0$ is the rest frequency and $c$ is the light speed.
The employed natural broadening parameter (Einstein A coefficient) $\Gamma_n$ was 2.55~$\times~10^8$~s\up{-1} and 2.56~$\times~10^8$~s\up{-1} for h and k line, respectively.
The Stark broadening was calculated as $\Gamma_s = 4.8\times10^{-7}\times~(n_e~[$cm\up{-3}])~s\up{-1} for both h and k lines. The Van der Waals damping is $\Gamma_{vw} = 6.6~\times~10^{-11} \times (T~[K])^{0.3} \times n_{\rm HI}$, where $n_{\rm HI}$ is the density of neutral hydrogen atoms \citep{mil74}.
Note that in these off-limb structures having rather low densities the main broadening is the Doppler one with a dominant micro-turbulent component in the Mg II lines.
Then, the total damping parameter is defined as $\Gamma=\Gamma_n+\Gamma_s+\Gamma_{vw}$ and the Voigt function takes the form 
\begin{equation}
H(a,x)=\frac{a}{\pi}\int_{-\infty}^{\infty}\frac{{\rm e}^{-y^2}}{(x-y)^2+a^2}{\rm d}y.
\end{equation}
Here, $a$ is the damping parameter given as $a=\Gamma/(4\pi\Delta\nu_D)$, $x$ is the dimensionless frequency offset given as $x=(\nu-\nu_0)/\Delta\nu_D$, and $y$ is defined as $y=\Delta\nu/\Delta\nu_D$.
The optical thickness as a function of wavelength can be expressed as 
\begin{equation}
\tilde{\tau}(x)=\tau_0\frac{H(a,x)}{H(a,0)},
\end{equation}
or
\begin{equation}
\tau(\Delta\nu)=\tilde{\tau}(x)|_{x=x(\Delta\nu)}.
\end{equation}

\par
The synthesized specific intensity (radiance) for the single-slab model ($I_{\rm s}$) was computed as
\begin{equation}
I_{\rm s}=\int_{t_\nu=0}^{t_\nu=\tau_{\nu}}S_\nu(t_\nu) {\rm e}^{-t_\nu}{\rm d} t_\nu,
\end{equation}
where $S_\nu$ is the line source function and $\tau_{\nu}$ is the optical thickness of the slab.

\begin{figure}[ht]
\centerline{\includegraphics[width=\linewidth]{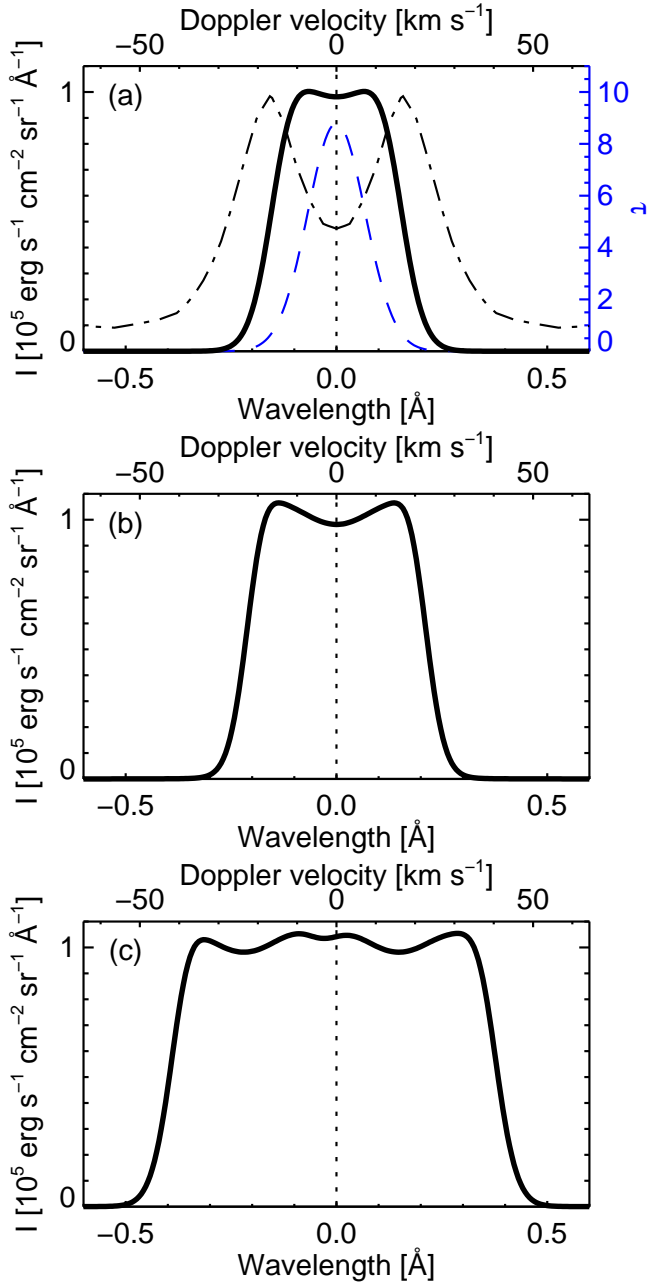}}
\caption{
An example of the Mg II k line profiles synthesized by (a) single-slab model, (b) multi-slab model without LOS velocities, and (c) multi-slab model with LOS velocities.
Details of the multi-slab model with randomly assigned LOS velocities and values of the LOS velocities are given in the caption of Figure~\ref{fig-step} and in the main text.
Input parameters of all models are $P$~=~0.1\pcgs, $T$~=~10$^4$\K, $D$~=~250\km, \Vturb~=~10\kms, and \Vradial~=~0\kms.
In panel (a), the specific intensity profile is shown in black solid line, the adopted illumination intensity profile at $H$ = 5000 km is shown in black dash-dotted line, and the optical thickness for each wavelength is shown in blue dashed line.
In panels (b) and (c), the specific intensity profile with 10 slabs along the LOS is shown.
In each panel, the axis indicating the LOS velocity values is shown at the top.
}
\label{fig-smr}
\end{figure}


\par
Figure~\ref{fig-smr}(a) shows an example of the synthetic Mg II k line profile from a single-slab model, with $P$~=~0.1\pcgs, $T$~=~10$^4$\K, $D$~=~250\km, \Vturb~=~10\kms, \Vradial~=~0\kms, which is a set of typical parameters of spicules.
With these parameters, the electron density at the center of the slab is about 2.7 $\times$ 10\up{10}~cm\up{-3}, the optical thickness at the line center is about 8.8, and the Mg II k line profile shows a little reversal, the peak intensity level of $\sim 1.0\times 10^5$ erg s\up{-1} cm\up{-2} sr\up{-1} \AA\up{-1}, and the line width measured by the 50\% bisector is 34\kms\ (0.32~\AA).
This line width is partially due to the opacity broadening and enough to explain the width of the observed profile at the upper part of the spicules -- for more details see discussion in Sect.~\ref{ss-high}.
Note that in the Figure~\ref{fig-smr}(a) we also show the illumination intensity profile at $H$ = 5000 km used here.

\par
Our choice of the typical geometrical width of spicules $D = $ 250\km\ is based on the work by \cite{per12}. 
These authors obtained the mean diameter of quiet-Sun and coronal hole spicules in Ca II H images as 250\km\ and 340\km, respectively, using space-borne observations by Hinode/SOT. 
Although in the models used in the present work we compute the ionization equilibrium of hydrogen to obtain the electron density used for non-LTE modeling of magnesium, we may assume that the observable geometrical width of spicules in Ca II H is similar to that in the H$\alpha$ line.
This is because the formation temperature of H$\alpha$ is similar to the formation temperature of Ca II H and the optical thickness of the H$\alpha$ line is also similar to the optical thickness of the Ca II H line. 
In fact, \cite{per13} show in their Figure 3 that both the structure of spicules and their time evolution are very similar in H$\alpha$ and Ca II H observed by Hinode/SOT.
However, we note that \cite{pas09} measured quiet-Sun spicule widths of around 660\km, using ground-based H$\alpha$ observations. 
These observations were performed in a different region and at a different time from those of \cite{per12}. 
Also, the H$\alpha$ observations by \cite{pas09} were obtained using a narrow-band filter while the Ca II H observations by \cite{per12} were obtained using a broad-band filter.
The difference between the filters may result in different apparent widths of the observed spicules.
Moreover, the seeing-affected ground-based observations might result in a smearing of the observed fine structures and thus to larger perceived widths of these structures. 
Therefore, we adopt the results of \cite{per12} in the present work. 
Additionally, the conclusions of the present paper would not be significantly affected, even if we would assume a large width (for example $D =$ 500\km) as the typical width of spicules.


\begin{figure*}[ht]
\centerline{\includegraphics[width=\linewidth]{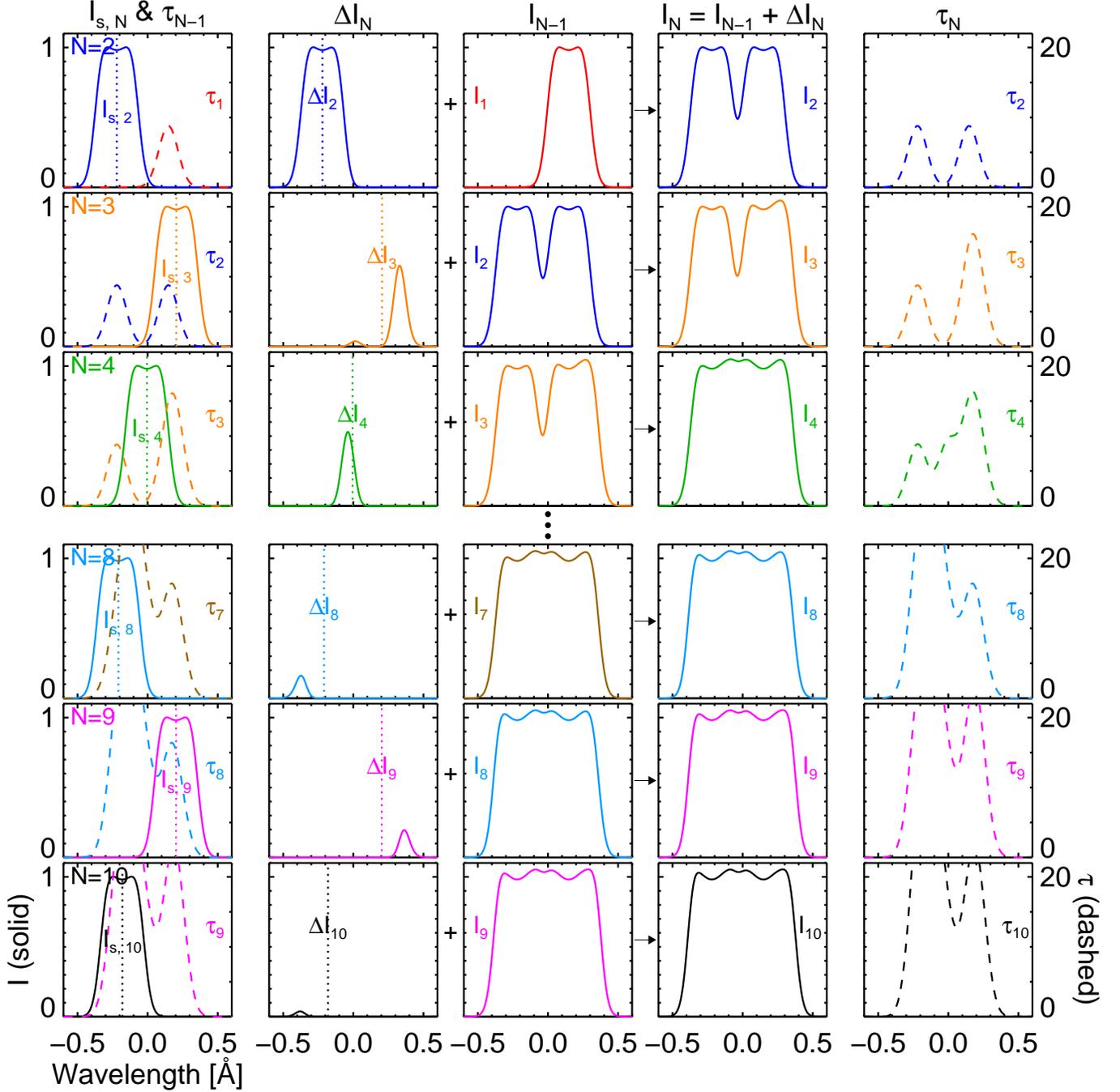}}
\caption{
Details of the multi-slab model with LOS velocities shown in Figure~\ref{fig-smr}(c).
The panel is divided into 6 rows, each of which corresponds to the model with $N$ slabs.
The first column shows the line profile of the $N$-th slab as a single slab ($I_{{\rm s}, N}$, solid line).
The total optical thickness of the ($N$$-$$1$) slabs that stands in front of the $N$-th slab is also shown ($\tau_{N-1}$, dashed line).
The second column shows the additional intensity $\Delta I_{N}=I_{{\rm s}, N}\exp(-\tau_{N-1})$, where the $N$-th slab is added and its intensity ($I_{{\rm s}, N}$) is attenuated by the ($N$$-$$1$) slabs with the total optical thickness of $\tau_{N-1}$ in front of the $N$-th slab.
The third and the fourth column show the total intensity profile before and after the $N$-th slab is added, i.e. $I_{N-1}$ and $I_{N}$, respectively.
The last column shows the total optical thickness with $N$ slabs ($\tau_N$). 
This attenuates the intensity from the ($N$$+$$1$)-th slab, if the ($N$$+$$1$)-th slab is added.
We use here the following randomly generated set of LOS velocities for slabs N=1 to N=10: 6, -24, 22, -1, -13, -14, -9, -22, 22, -19 \kms.
These velocities are indicated in the first and second columns by the vertical dotted lines.
The horizontal and vertical axes are common for all panels.
The left vertical axis in the first column is for the specific intensities with the scaling 10$^5$ erg s\up{-1} cm\up{-2} sr\up{-1} \AA\up{-1} and right vertical axis in the last column is for the optical thicknesses.
The horizontal axis is for the wavelength scale with respect to the line center of the Mg II k in the rest frame.
}
\label{fig-step}
\end{figure*}

\subsection{Multi-slab Model without LOS Velocities} \label{ss-multi}
In this model, we experimented with placing additional static slabs along the LOS to study how the synthetic Mg II profiles change when more than one slab is assumed.
The multi-slab model here consists of a set of $N$ identical 1D slabs along the LOS (Figure~\ref{fig-model}(b)).
In this subsection, we assume static slabs to firstly investigate the superposition effect, although we acknowledge that this situation is not realistic (for more realistic case, see Sec.~\ref{ss-random}).
Using the synthetic intensity of the single-slab model $I_{\rm s} (\Delta\lambda)$ described in the previous section, the intensity of the multi-slab model without LOS velocities $I_{\rm m} (\Delta\lambda)$ is obtained as
\begin{equation}
I_{\rm m}(\Delta\lambda)=I_{\rm s}(\Delta\lambda)\sum_{i=1}^{N}\exp\{-(i-1)\tau(\Delta\lambda)\},
\end{equation}
where the intensity from the {\it i}-th slab is attenuated by the slabs that stand in front of it.

\par
Figure~\ref{fig-smr}(b) shows an example of the synthetic Mg II k line profile obtained by multi-slab modeling using a set of $N$ identical slabs. 
Each slab assumes the same set of input parameters as the single-slab model described in Sect.~\ref{ss-single}.
The number of slabs varies from 1 to 10 and the result with $N=1$ is the same as shown in Figure~\ref{fig-smr}(a).
As we add additional slabs, the resulting intensity profiles get enhanced at the wavelengths in which the total optical thickness is less than unity, i.e. we can see more spicules along the LOS in the line wings.
The width of the synthetic profile produced by static multi-slab model does not increase after a certain number of slabs is added.
In the example shown in Figure~\ref{fig-smr}(b), the width of the resulting profile becomes saturated after adding less than 10 slabs.
The line width measured by the 50\% bisector method for the case of $N$ = 10 is 46\kms\ (0.43~\AA).
Such a width is not sufficient to explain the observed line widths at lower and middle altitudes, where the line width measured by the same method is up to 110\kms ($\sim$1.0~\AA).
For more details see Sect.~\ref{ss-mid}.


\subsection{Multi-slab model with Random LOS Velocities} \label{ss-random}
In this subsection, we simulate more realistic situation with multiple slabs to which we randomly assign LOS velocities (Figure~\ref{fig-model}(c)).
The origin of the LOS velocities used in this model is following.
Because spicules are typically inclined with respect to the solar surface (an inclination of 20\up{\circ} to 37\up{\circ} is given by \citealt{tsi12}), we need to take into account motions both parallel and perpendicular to the spicule axis.
The typical amplitude of transverse velocities (i.e. perpendicular to the spicule axis) is $\sim$~25\kms \citep{dep07}.
Typical apparent upward velocities (i.e. parallel to the spicule axis) are of the order of 25\kms\ on the quiet Sun \citep{pas09}. 
We assume here that the same is true in a coronal hole.
Moreover, from our analysis of the IRIS observations presented in Sect.~\ref{ss-analysis} we obtained in the upper parts of spicules typical unsigned LOS velocities up to 25\kms.
Therefore, in the present study, we adopted a LOS velocity that is randomly selected from a uniform distribution ranging from -25 to 25\kms.

\par
In the case of a multi-slab model with randomly assigned LOS velocities, the specific intensity $I_{\rm r}(\Delta\lambda)$ is obtained as
\begin{equation}
\begin{split}
I_{\rm r}(\Delta\lambda) 
& =I_{\rm s}(\Delta\lambda-\Delta\lambda_{V_{{\rm LOS}, 1}})\\
& +\sum_{i=2}^{N} \Bigl[\Bigr. I_{\rm s}(\Delta\lambda-\Delta\lambda_{V_{{\rm LOS}, i}}) \\
& \times\exp\bigl\{-\sum_{l=1}^{i-1}\tau(\Delta\lambda-\Delta\lambda_{V_{{\rm LOS}, l}})\bigr\} \Bigl.\Bigr], \\
\end{split}
\end{equation}
where $N\geqq2$ and $\Delta\lambda_{V_{{\rm LOS}, i}}$ is the Doppler shift in the wavelength due to the LOS velocity of the {\it i}-th slab (see Figure~\ref{fig-model}(c)).
Similar scheme was applied in \citet{gun08} for 2D multi-thread modeling of prominence fine structures.
Due to the LOS velocity, the entire intensity profile emerging from {\it i}-th slab and the absorption profile of the {\it i}-th slab are shifted by $\Delta\lambda_{V_{{\rm LOS}, i}}$ with respect to the rest wavelength.
Such a Doppler-shifted intensity profile from {\it i}-th slab $I_{\rm s}(\Delta\lambda-\Delta\lambda_{V_{{\rm LOS}, i}})$ is then attenuated by all foreground slabs with the optical thickness $\Sigma_{l=1}^{i-1}\tau(\Delta\lambda-\Delta\lambda_{V_{{\rm LOS}, i}})$ as $I_{\rm s}(\Delta\lambda-\Delta\lambda_{V_{{\rm LOS}, i}})\exp\{-\Sigma_{l=1}^{i-1}\tau(\Delta\lambda-\Delta\lambda_{V_{{\rm LOS}, i}})\}$.

\par
Figure~\ref{fig-smr}(c) shows, as an example, the result of the multi-slab modeling with LOS velocities using a set of $N (= 10)$ identical slabs modeled with the MALI code. 
The details are shown in Figure~\ref{fig-step} and its caption.
For individual slabs we again use the same set of input parameters as in Sect.~\ref{ss-single}.
The number of slabs $N$ varies from 1 to 10.
The LOS velocities of the 1st slab to 10th slab used are 16, -24, 22, -1, -13, -14, -9, -22, 22, -19 \kms.
As we add additional slabs, the resulting intensity profiles get enhanced at the wavelengths in which the total optical thickness $\Sigma_{i=1}^{N}\exp\{-(i-1)\tau(\Delta\lambda)\}$ is less than unity, which is the same mechanism as demonstrated in the multi-slab modeling without LOS velocities in the previous subsection.
However, the line width in the case of $N$ = 10 is significantly larger (83\kms\ or 0.78~\AA) compared to the case of $N$ = 10 without LOS velocities (46\kms\ or 0.43~\AA) shown in Figure~\ref{fig-smr}(b).
This is due to the relative shifts of the intensity profiles and optical thickness profiles for each slab. 
Because the optical thickness in the wing of each profile is small, the intensity from the added slab can appear there if the added slab has the Doppler velocity that corresponds to the wing wavelength.
The line width in this case is dependent on the minimum and maximum LOS velocities, which are -24\kms\ and 22\kms\ in the 2nd and 9th slab, respectively.
This can help to explain the broad and flat profiles observed at middle heights, as we discuss in Sect.~\ref{ss-mid}.

\subsection{IRIS Instrumental Profile} \label{ss-convol}
The IRIS NUV instrumental profile has the FWHM of $\Delta\lambda_{\rm inst} = $ 50.54~m\AA.
This instrumental broadening affects the observed line profile and thus needs to be considered also for the synthetic profiles.
We take the instrumental broadening into account in the following way:
\begin{equation}
\begin{split}
I(\Delta\lambda)
&=\frac{1}{\sqrt{2\pi\sigma_{\rm inst}}}\times \\
&\int I_{synt}(\Delta\lambda^\prime) \exp\left[-\frac{(\Delta\lambda-\Delta\lambda^\prime)^2}{2\sigma_{\rm inst}^2}\right] d(\Delta\lambda^{\prime}),
\end{split}
\end{equation}
where $\sigma_{\rm inst} = \Delta\lambda_{\rm inst} / (2\sqrt{2\ln2})$.
This convolution is applied to all synthetic profiles obtained by the single-slab model, multi-slab model without LOS velocities, and multi-slab model with random LOS velocities, which are presented in this work.

\section{PARAMETER DEPENDENCE OF THE MG II H AND K LINE PROFILES} \label{s-para}
\begin{figure*}[t]
\centerline{\includegraphics[width=0.98\linewidth]{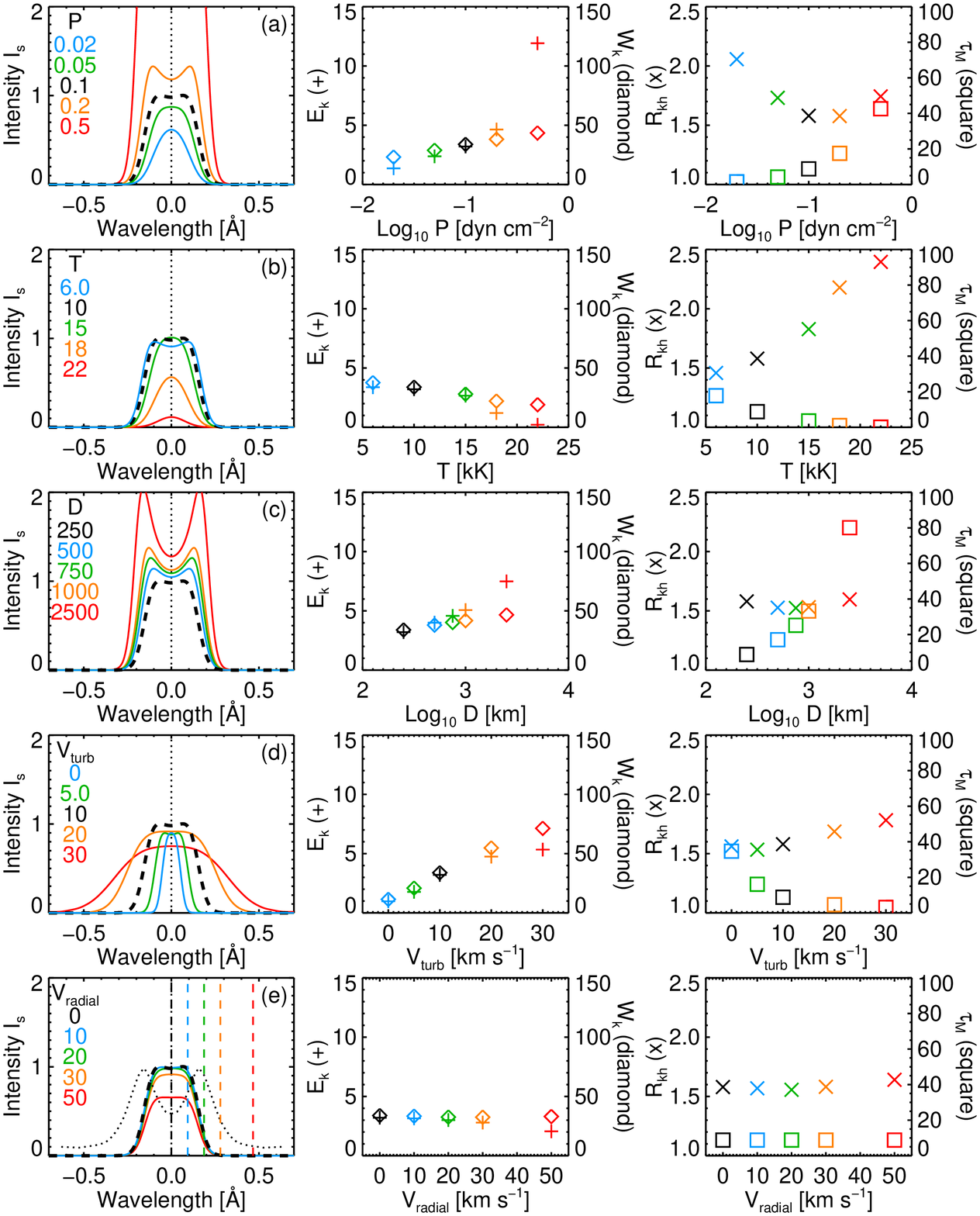}}
\caption{
Parameter dependence of the synthetic Mg II k profiles produced by single-slab models.
The left column shows specific intensity profiles $I_s$ (in $10^5~\times$ \intensity).
The parameter that is varied in each panel of the left column is (a) pressure $P$ (in \pcgs), (b) temperature $T$ (in kK), (c) thickness of the slab $D$ (in km), (d) micro-turbulent velocity \Vturb\ (in \kms), and (e) radial velocity \Vradial\ (in \kms).
In panel (e), the illumination profile is shown in the dotted line and the velocities of adopted radial velocities are shown in vertical dashed lines.
The black dashed profiles represent models with the common set of parameters ($P$ = 0.1\pcgs, $T$ = 10~kK, $D$ = 250\km, \Vturb\ = 10\kms, and \Vradial\ = 0\kms) and are thus identical in all panels.
The profiles in blue, green, orange, and red solid lines represent models with the values of the corresponding parameter ranging from smaller to larger.
The colors of the profiles correspond to the colors of the inlaid numbers. 
In the middle column, integrated intensities \Ek\ (in 10$^4$ \intensity) and the widths (\Wk\ in \kms) for the models in the left column are shown.
In the right column, the ratios of integrated intensities in the Mg II k and h lines \Rkh (= \Ek /\Eh) and the maximum optical thicknesses ($\tau_M$) are shown.
}
\label{fig-para-s}
\end{figure*}

\begin{figure*}[!ht]
\centerline{\includegraphics[width=0.98\linewidth]{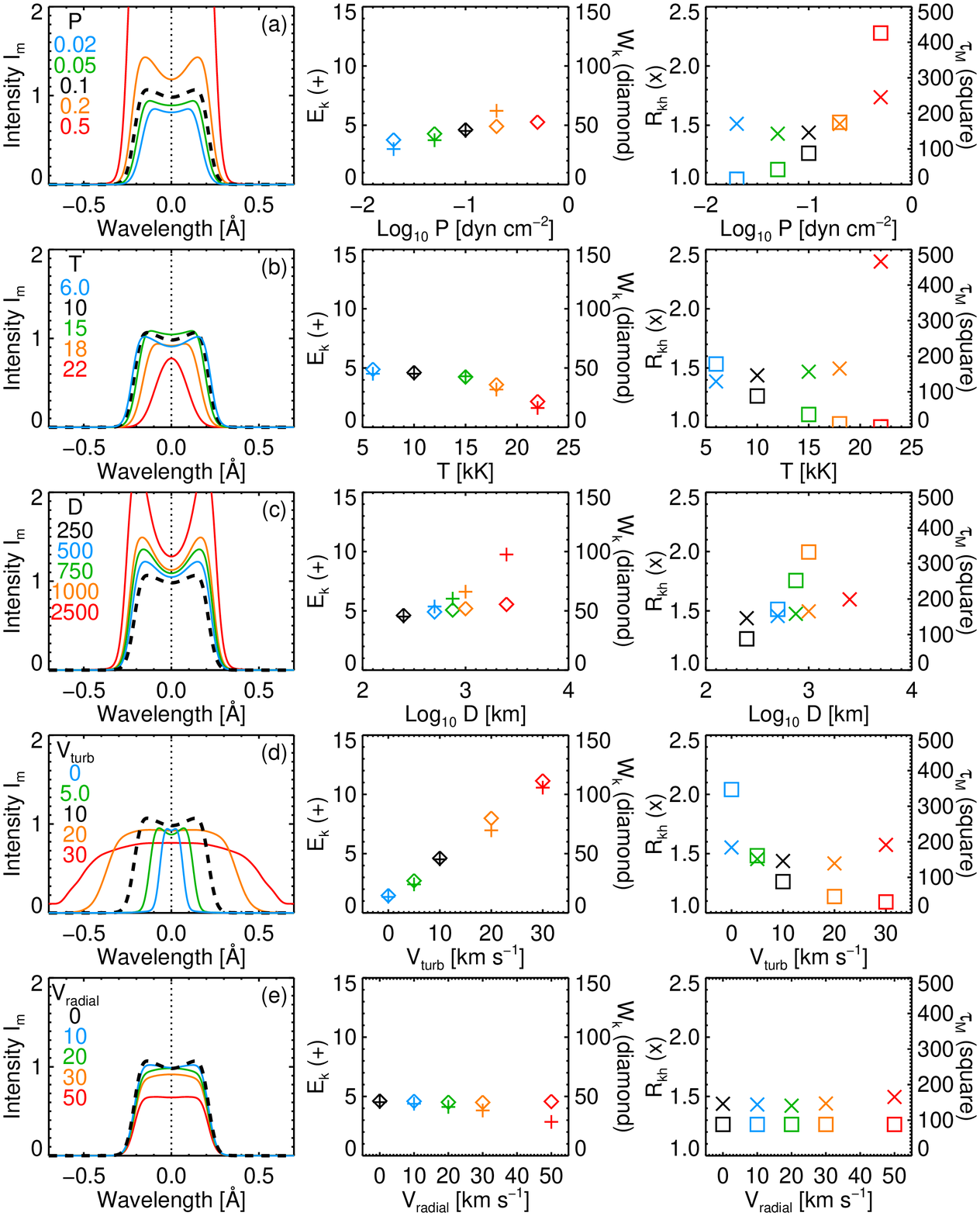}}
\caption{
The same as in Figure~\ref{fig-para-s} but for multi-slab models without LOS velocities.
}
\label{fig-para-m}
\end{figure*}

\begin{figure*}[ht]
\centerline{\includegraphics[width=0.98\linewidth]{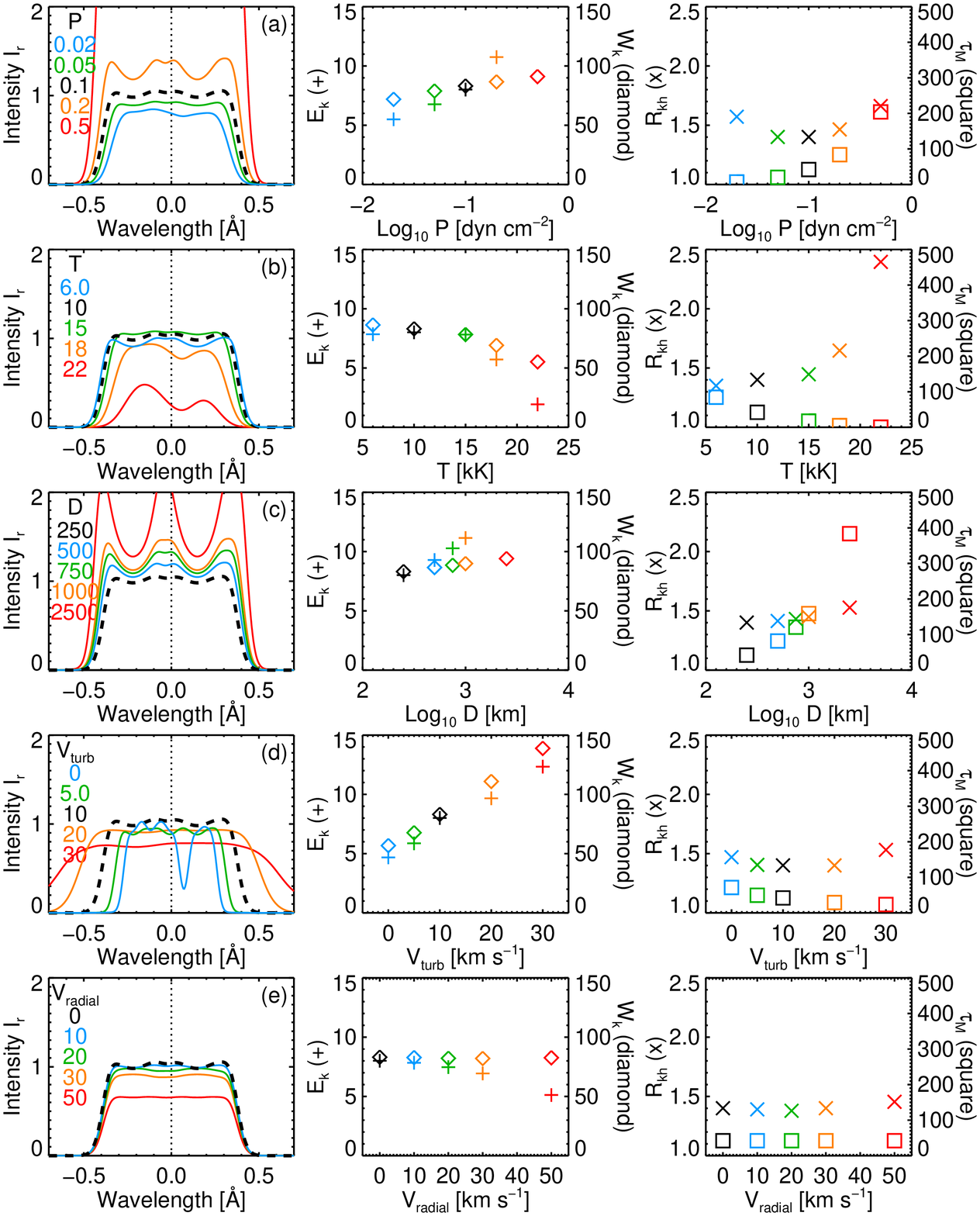}}
\caption{
The same as in Figure~\ref{fig-para-s} but for multi-slab models with randomly assigned LOS velocities.
}
\label{fig-para-r}
\end{figure*}

We computed a grid of single-slab and multi-slab models (with and without LOS velocities) for a set of different input parameters and for each model we synthesized the Mg II h and k line profiles.
In Figures~\ref{fig-para-s}, \ref{fig-para-m}, and \ref{fig-para-r}, we show an example of how the Mg II k line profile is modified when we change one of the input parameters, i.e. the pressure $P$, temperature $T$, geometrical thickness of the single-slab $D$, micro-turbulent velocity \Vturb, or radial velocity \Vradial, in different models: single-slab model, multi-slab model without LOS velocities, and multi-slab model with LOS velocities, respectively.
We used $P$ = 0.1\pcgs, $T$ = 10$^4$\K, $D$ = 250\km, \Vturb\ = 10\kms, and \Vradial\ = 0\kms\ as a common set of parameters and change one of them in each panel of Figures~\ref{fig-para-s}, \ref{fig-para-m}, and \ref{fig-para-r}.
All synthetic profiles are shown after the convolution with the instrumental broadening.
Each panel of Figures~\ref{fig-para-s}, \ref{fig-para-m}, and \ref{fig-para-r} includes additional two plots, which indicate derived values of integrated line intensity of Mg II k line (\Ek), the line width of Mg II k line obtained by the 50\% bisector method (\Wk), the ratio of the integrated line intensities of Mg II k and h lines \Rkh (= \Ek /\Eh), and the maximum optical thickness ($\tau_M$) as a function of the changing parameters.
The colors used in the inlaid text correspond to the colors of the plotted profiles and plotted symbols.

\subsection{Single-slab model} \label{ss-para-s}
Synthetic Mg II k profiles produced by the single-slab model are shown in Figure~\ref{fig-para-s}. 
In Figure~\ref{fig-para-s}(a), we show the dependence on the different values of the pressure (the used pressure values are indicated in the figure).
From the plotted synthetic profiles, we can clearly see that the synthetic profiles become very intense with increasing pressure. 
The optical thickness $\tau_M$ increases and is almost proportional to the pressure.
The integrated intensities \Ek\ also increase with the pressure.
The ratio of the integrated intensities of the h and k lines is large and more than 2 for the lowest pressure value (0.02 \pcgs).
Line widths become larger with increasing pressure but they do not exceed  50\kms ($\sim$ 0.5~\AA) even for very high pressure of 0.5\pcgs.

\par 
In Figure~\ref{fig-para-s}(b), we demonstrate the dependence of the synthetic Mg II k profiles on the temperature. 
The intensity steeply decreases when the temperature reaches above 18~kK. 
Optical thickness $\tau_M$ also decreases with temperature.
This is because of the ionization of the Mg II ion to Mg III.
With the temperature of 22~kK, the most Mg II ions are ionized into Mg III \citep{hei14} and the modeled spicule becomes optically thin ($\tau_M\sim$ 0.1) in the Mg II k line.
In the temperature range $T >$ 18~kK, the number of Mg II ions decreases, the integrated intensity \Ek\ becomes significantly smaller, and the line intensity ratio \Rkh\ becomes somewhat larger than two.
The line widths \Wk\ are always less than 40\kms\ ($\sim$ 0.4~\AA) and are not too sensitive to the temperature and decrease slightly (from 38 to 19\kms) between temperature of 6~kK and 22~kK. 

\par 
The Figure~\ref{fig-para-s}(c) shows how the Mg II k profiles change when we modify the geometrical thickness of the modeled slab, from 250\km\ to 2500\km.
The optical thickness is nearly proportional to the geometrical thickness.
The Mg II k line is more and more reversed for larger geometrical thicknesses.
This is because with the increasing geometrical (and thus also the optical) thickness, the layer of $\tau \sim$ 1 in the line core is located closer and closer to the surface where the source function decreases.
On the other hand, the line widths \Wk\ do not change significantly and stay below 50\kms\ ($\sim$ 0.5~\AA).
Although the integrated intensity \Ek\ increases with the geometrical thickness, the ratio of integrated intensities of h and k lines \Rkh\ is nearly constant.

\par
In Figure~\ref{fig-para-s}(d), we change the micro-turbulent velocity.
The integrated intensity \Ek\ increases with the micro-turbulent velocity but practically saturates above \Vturb\ of 20\kms.
The line width \Wk\ increases dramatically with the micro-turbulent velocity and the resulting Mg II line profiles have significantly broadened wings.
As the micro-turbulent velocity increases, the line ratio \Rkh\ increases because the line profile broadens and the optical thickness at the line center $\tau_M$ decreases.

\par 
In Figure~\ref{fig-para-s}(e), we show dependence of synthetic Mg II profiles on the upward (radial) velocity, which causes the Doppler dimming/brightening effect.
In this case, the plasma properties such as temperature, pressure, density do not change and thus the optical thickness remains the same.
However, the peak intensity changes with the change of the upward velocity.
This is because the intensity at each wavelength changes due to the Doppler shifts of the illumination profile (indicated in dotted curve in the left panel of this row).
We show here for the first time how the Mg II h and k line intensity of spicules is sensitive to the radial velocity via the Doppler brightening/dimming effects.
\cite{zha12} reported that typical upward velocities of spicule tops are about 13\kms\ in a quiet region and about 38\kms\ in a coronal hole.
We demonstrate here that upward velocities above 30\kms\ cause Doppler dimming.

\subsection{Multi-slab model without LOS velocities} \label{ss-para-m}
In Figure~\ref{fig-para-m}, we show synthetic Mg II k profiles obtained by multi-slab model without LOS velocities (for more details see Sect.~\ref{ss-multi}).
Because we assume 10 identical slabs, the optical thickness for each set of model input parameters is 10 times larger than that of corresponding single-slab model.  
Therefore, in almost all cases, the optical thickness at the line center is very large and the Mg II k line has reversed profiles.
Due to the large total optical thickness of multiple slabs, line widths \Wk\ for each set of input parameters are only slightly larger than in the case of a single-slab model and do not exceed 60\kms (0.56~\AA).
The exceptions are the very broad profiles with large values of micro-turbulent velocities.
The integrated line intensities \Ek\ are mostly less than twice as intense as those from corresponding single-slab models even though we use 10 identical slabs.
This is again caused by the large optical thickness of individual slabs.
Exceptions are the cases of the largest temperatures of $T$ = 18~kK and 20~kK in which, however, the optical thickness is small. 
The ratio of the integrated intensities in the k and h lines \Rkh\ is around 1.3--1.5 except in the very high temperature case with $T$ = 2.2~kK, in which the Mg II k line has optical thickness around one.

\subsection{Multi-slab model with LOS velocities} \label{ss-para-r}
Results of the multi-slab model with random LOS velocities (see Sect.~\ref{ss-random}) are shown in Figure~\ref{fig-para-r}.
The maximum optical thickness $\tau_M$ for each set of input parameters is significantly lower (about half) than that of the corresponding multi-slab model without LOS velocities.
This is because now each slab has its own LOS velocity and both the intensity and optical thickness profiles of individual slabs are mutually Doppler-shifted.
Even so, the Mg II k line is optically thick for all sets of input parameters except the set with the highest temperature ($T$ = 22~kK).
The line width \Wk\ is now about 1.5 times larger compared to the corresponding multi-slab model without LOS velocities and typically reaches values between 70\kms\ (0.65~\AA) and  90\kms\ (0.84~\AA).
Such large widths are more than twice the widths produced by the single-slab model and cannot be achieved by either single-slab or multi-slab model without LOS velocities unless we use very large micro-turbulent velocities.
We note that line widths between 80 and 120\kms\ are typically obtained by the bisector method from the observations at middle or lower altitudes around $Y$ = 0\arcsec--10\arcsec\ analyzed here (see Figure~\ref{fig-map}(d), \ref{fig-dist}(d)).
The integrated intensity \Ek\ is also about 1.5 times larger than that in multi-slab models without LOS velocities but the ratio of integrated intensities \Rkh\ is almost the same.
The exceptions are the input parameter sets with the highest temperatures (18~kK and 22~kK).
Note that in Figure~\ref{fig-para-r} we show results with identical radial velocities applied to all slabs in the multi-slab model with LOS velocities. 
We acknowledge that such an assumption may not be entirely realistic. 

\subsection{Ratio of integrated intensity of Mg II h and k lines} \label{ss-para-ratio}
The ratio of the integrated intensities of Mg h and k lines is sensitive to the optical thickness and can be larger than 2 only when the lines have optical thickness around unity or below.
From the set of the single-slab models that we show here, this is the case only for models with the smallest pressure (0.02\pcgs) or the highest temperatures (18~kK or 22~kK).
For the multi-slab models, it is only the model with the highest temperature of 22~kK.
We note that higher values of \Rkh\ are identified only at higher altitudes in the observed data set studied here (see Figure~\ref{fig-map}(b)).
At these altitudes ($Y$ $>$ 1\arcsec), majority of the obtained \Rkh\ values is below 1.8. 
However, a fraction of the observed data exhibits \Rkh\ values around 2 or higher (see Figure~\ref{fig-dist}(b)). 

\begin{figure*}[ht]
\centerline{\includegraphics[width=1.0\linewidth]{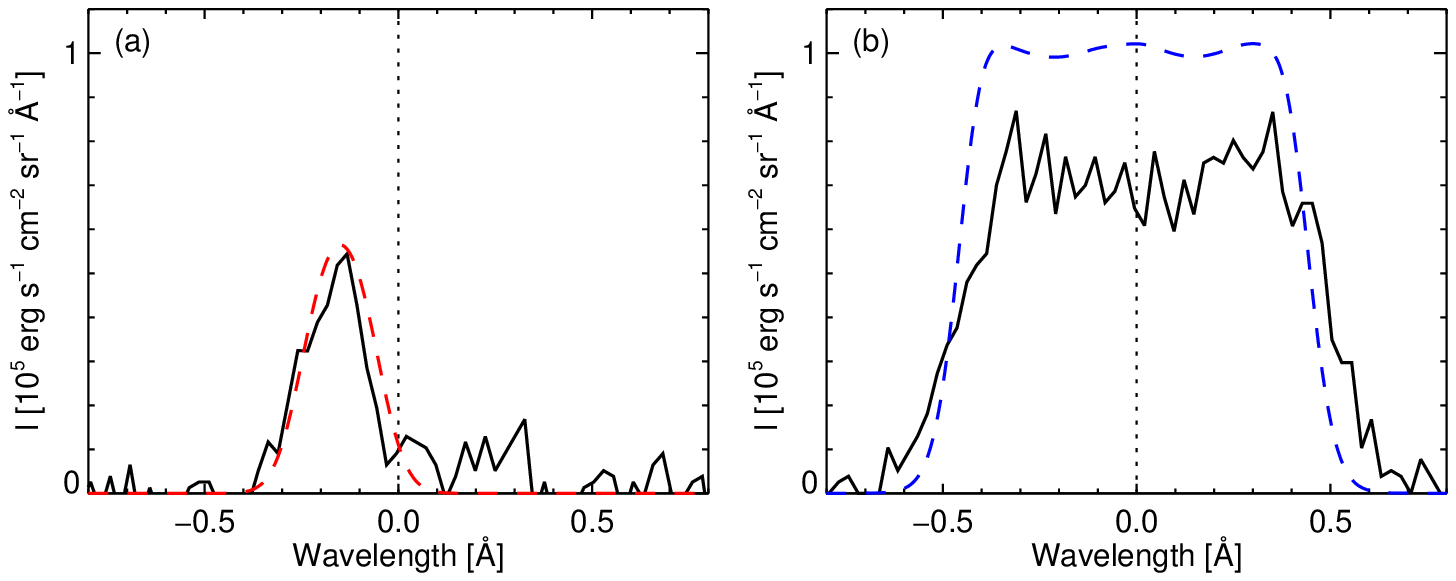}}
\caption{
Comparison of the observed and modeled Mg II k line profiles.
(a)
The black solid line is the observed profile at $Y$ = 15.0\arcsec\ and time step 324.
The red dashed line is the profile produced by single-slab model with parameters $P$ = 0.1\pcgs, $T$ = 18~kK, $D$ = 250\km, \Vturb\ = 10\kms, and \Vradial\ = 0\kms.
The synthetic profile is shifted by 0.15~\AA\ ($\sim$ 16\kms) towards shorter wavelengths to be comparable with the observed profile.
(b)
The black solid line is the observed profile at $Y$ = 5.0\arcsec\ and time step 350.
The blue dashed line shows the profile produced by multi-slab model with random LOS velocities with 10 identical slabs with parameters $P$ = 0.1\pcgs, $T$ = 10~kK, $D$ = 250\km, \Vturb\ = 15\kms, and \Vradial\ = 0\kms.
}
\label{fig-comp}
\end{figure*}

\section{DISCUSSION} \label{s-discussion}
\subsection{Mg II h and k line profiles at higher altitudes} \label{ss-high}
Our analysis of IRIS Mg II observations presented in Sect.~\ref{s-obs} shows that at higher altitudes ($Y \sim$ 15\arcsec) Mg II h and k profiles are relatively narrow and generally exhibit lower integrated intensities than profiles obtained at lower altitudes.
Typical widths of the Mg II k profiles \Wk\ at higher altitudes are a few tens of \kms\ (see Figures~\ref{fig-map}(d) and \ref{fig-dist}(d)) and the integrated line intensities \Ek\ reach up to a few $\times 10^4$ \intensity\ (see Figures~\ref{fig-map}(a) and \ref{fig-dist}(a)). 
The ratio of integrated intensities of h and k lines (\Rkh) reaches up to 2.0 (Figures~\ref{fig-map}(b) and \ref{fig-dist}(b)).
The observed profiles at higher altitudes are often strongly Doppler-shifted and exhibit unsigned LOS velocities up to 20\kms\ (Figure~\ref{fig-map}(c) and \ref{fig-dist}(c)). 
In the present work, we argue that such Mg II profiles represent an emission of individual spicules that reach above the spicular forest. 
To support this argument, we show here that these profiles can be reproduced by narrow single-slab models representing a single spicule. 
In Figure~\ref{fig-comp}(a), we show an example of a typical Mg II k line profile obtained at the higher altitude -- in this particular case at altitude $Y$ = 15\arcsec\ at the time step 324. 
The width of this profile \Wk\ is 21\kms, the integrated line intensity \Ek\ = 1.5 $\times 10^4$ \intensity, the ratio of integrated intensities \Rkh\ = 1.9, and the line shift \VLOSk\ $= -19$\kms.
In the present work, we do not aim to find a model with the best fit to this observed profile. 
Rather, from the set of profiles shown in Figures~\ref{fig-para-s} and \ref{fig-para-r}, we have selected one profile that closely resembles the observed profile. 
In Figure~\ref{fig-comp}(a), we plot the synthetic profile (the red dashed line) red-shifted by 0.15~\AA\ ($\sim$16\kms). 
This synthetic profile corresponds to a single-slab model with $P$ = 0.1\pcgs, $T$ = 18~kK, $D$ = 250\km, \Vturb\ =10\kms, and \Vradial\ = 0\kms.
The used geometrical thickness (250 km) is consistent with the observations \citep[e.g.][]{per12}.
The profile has the width \Wk\ = 22\kms, \Ek\ = 1.2 $\times 10^4$ \intensity, and \Rkh\ = 2.2.
These profile parameters are quantitatively comparable with the parameters of the selected observed profile. 

\par
When we assume the geometrical thickness of a spicule to be 250~km, only optically thin slab with high temperature or low pressure plasma can achieve the observed single-peaked profile with small integrated intensity (see Figure~\ref{fig-para-s}).
In our observations, a fraction of the upper part of spicules has large values of the ratio of the k and h lines \Rkh\ (around 2 or higher).
Such large line ratios can be achieved only when the line is optically thin due to its high temperature ($\sim$20~kK) or low pressure ($\lesssim$~0.02\pcgs) as we argue in Sect.~\ref{ss-para-ratio}.
A pressure of 0.02\pcgs\ represents a rather low, coronal value that might be too low for spicules in the static case.
However, in a more realistic dynamic situation, cooling due to an expansion might occur in the upper parts of spicules. 
Such a scenario could lead to rather low pressure values. 
On the other hand, the combination of pressure of 0.1\pcgs\ (typical in spicules) and temperature of 20~kK (typical chromospheric values) might be more realistic.
Moreover, recent imaging observations suggest that upper parts of spicules are visible in the higher temperature, transition region lines \citep{per14}.
If optically thin plasma is caused by high temperature ($\sim$20~kK), one possible heating mechanism could be a shock heating.
In such a case, the possible causes of the shocks in relation to spicule dynamics needs to be investigated.
To gain better understanding of why and how such high temperature or low pressure plasma is produced, we will need to follow individual spicules throughout their lifetimes with multi-line spectroscopy and high-resolution imaging.

\subsection{Mg II h and k line profiles at middle altitudes} \label{ss-mid}
The observations at the middle altitudes ($Y \sim$ 5\arcsec) show Mg II profiles that are significantly broader than profiles obtained at higher altitudes. 
Figures~\ref{fig-map}(d) and \ref{fig-dist}(d) show line widths \Wk\ between 80 and 110\kms. 
The integrated intensities \Ek\ are around 8 $\times 10^4$ \intensity\ and the ratio of integrated intensities \Rkh\ is 1.2-1.4 (see Figures~\ref{fig-map}(a-b) and \ref{fig-dist}(a-b)). 
The LOS velocities derived from the observed Doppler shifts \VLOSk\ are between $-5$ and $+5$\kms\ (Figures~\ref{fig-map}(c) and \ref{fig-dist}(c)), which are significantly lower than the LOS velocities derived at higher altitudes. 
In Figure~\ref{fig-comp}(b), we show an example of a typical Mg II k line profile obtained at the altitude $Y$ = 5\arcsec\ at the time step 350. 
The parameters of this profile are \Wk\ = 100\kms, \Ek\ = 7.6 $\times$ 10$^4$ \intensity, and \Rkh\ = 1.3. 
The selected profile is flat-topped and shows multiple small intensity variations at the top. 

\par
To reproduce the observed Mg II k profile shown in Figure~\ref{fig-comp}(b), we selected a model from the grid of models shown in Figure~\ref{fig-para-r}. 
We again note that we do not aim here to find a model with the best fit to the observation. 
The selected model is shown in blue dashed line in Figure~\ref{fig-comp}(b), which represents the multi-slab model with randomly assigned LOS velocities from interval of $-25$ to $+25$\kms, $P$ = 0.1\pcgs, $T$ = 10~kK, turbulent velocity \Vturb\ = 15\kms, \Vradial\ = 0\kms, and 10 identical threads with a width $D$ = 250 \km.
This model produces a synthetic Mg II k profile with parameters \Wk\ = 98\kms, \Ek\ = 9.3 $\times$ 10$^4$ \intensity, and \Rkh\ = 1.4 that are quantitatively comparable to the observed profile.
The used LOS velocities with the maximum unsigned value of 25\kms\ are consistent with the Hinode Ca II H observations of the apparent velocity amplitude reported by \cite{dep07}.
In addition, the maximum line shifts at upper altitudes in our spectroscopic observations are also about 25\kms\ (see Figure~\ref{fig-dist}(c)).
Therefore, we suggest that the apparent velocities in the Hinode Ca II H imaging data correspond to the chromospheric gas motions in spicules and that velocities in the LOS direction are typically between $\pm$ 25\kms.
This multi-slab modeling with LOS velocities succeeds in producing broad Mg II k profiles without using the large \Vturb\ values like 25\kms.
We note that such a large \Vturb\ is above the local sound speed and thus may not be entirely realistic. 
The multi-slab models with random LOS velocities achieve broad Mg II h and k profiles by Doppler-shifting the intensity and optical thickness profiles of individual slabs with respect to each other. 
As we demonstrate in Figure~\ref{fig-step}, such mutual shifts lead to significantly broad profiles even when the profiles produced by individual slabs are narrow.
Therefore, each individual slab can have turbulent velocity below the local sound speed. 
Moreover, the synthetic profiles produced by multi-slab models with LOS velocities have an added advantage that they exhibit flat tops with a certain level of fine structuring, albeit not as pronounced as in the observed profiles.
The intensity level at the top of the modeled profile is slightly higher than the observed profile.
However, the intensity level of the modeled profile would naturally decrease if we considered the doppler dimming/brightening effects for each spicule slab depending on the relative values between the velocities of illuminating materials and the velocities of the slabs.
Another possibility is that if the spicule would be modeled as a vertical cylinder, it would emit radiation in all directions and the source function would thus decrease. 
Then the line-core emission, which is roughly equal to the source function, will be lowered \citep{hea77}. 

\par
We note that the broad observed Mg II h and k profiles, such as the one shown in Figure~\ref{fig-comp}(b) cannot be reproduced just by increasing the temperature in the models. 
This is because the Mg II ion is heavy and the Mg II h and k lines are thus not sensitive to the thermal broadening. 
In fact, the thermal width of the Mg II k line is only 2\kms\ (0.02~\AA) for the temperature of 6~kK and 4\kms\ (0.04~\AA) for 22~kK. 
When we compare this to the turbulent broadening of 10\kms (0.09~\AA) due to \Vturb, we can see that the Mg II h and k line widths observed at middle altitudes must be caused by some form of dynamics. 
One can either consider a large micro-turbulent velocity such as 25\kms, or assume macroscopic LOS velocities of multiple components along a LOS.

\par
Recently, \citet{ali18} have investigated the Mg II h and k line profiles observed by IRIS in a quiet-Sun polar region.
To understand the averaged line profiles at all altitudes above the limb, these authors conducted one-dimensional single-slab modeling with the fixed thickness of 500\km\ to synthesize the spectra and derived physical quantities for different heights by comparing temporally averaged observed spectra and synthetic spectra.
To achieve a good fit, \citet{ali18} considered a large turbulent velocity exceeding 20 \kms.
However, it can be expected that the Mg II h and k line profiles at lower part of chromosphere are a product of a superposition of several spicules along the LOS.
As we have shown in the present work, in case of such a superposition of spicules, the large turbulent velocity is not necessary to achieve the broad observed profiles if individual spicules have different LOS velocities.

\subsection{Mg II h and k line profiles near the limb} \label{ss-low}
At altitudes near the limb ($Y \sim$ 2\arcsec), the observed Mg II profiles are broad and usually distinctly double-peaked with a deep central reversal (see the example in Figure~\ref{fig-irissp}(b)). 
The intensity of the peaks varies significantly over time as can be seen in the online movie (link here). 
The 1D models in any of the configurations used in the present work are not able to reproduce such a type of profiles. 
To synthesize these profiles we need to include more complex geometry into our models of spicules. 
We will investigate these profiles in a future study. 

\section{CONCLUSIONS} \label{s-conclusion}
In the present paper, we use an extended, short exposure time (5.4~s) set of the Mg II h and k spectra of spicules in a polar coronal hole obtained by IRIS.
We quantitatively analyzed this set of observations to study the measured line intensities and their ratios, the line widths, and the derived LOS velocities as a function of time and altitude.

\par 
From this analysis, we found that the largest unsigned LOS velocities are mostly located in the highest part of the observed spicules.
Figure~\ref{fig-dist}(c) shows that the averaged unsigned LOS velocity is around 10\kms\ at high altitudes ($Y \sim$ 15\arcsec) while a large number of spicules exhibits LOS velocities around 20 \kms.
In the upper part of the observed spectra, we can also clearly identify very tall individual spicules, such as the strongly blue-shifted event visible in the observed spectra (Figure~\ref{fig-map}(c)) at around time 1750~s.
Such spicules would be easy to identify in high-resolution imaging observations.
The observed data thus suggest that in the upper part of spicules we can derive realistic information about the dynamics of a few or individual spicules.
In addition, we found that the line widths are the narrowest at the upper part.
For example, the averaged Mg II k line width derived by the 50\% bisector method is around 25\kms\ (0.23~\AA) at $Y \sim$ 15\arcsec\ (see Figure~\ref{fig-dist}(d)).
Such narrow line widths again suggest that in the upper part of the observed spectra we observe a few or even individual spicules.
Moreover, the ratio of integrated intensities of Mg II k and h lines (\Rkh) at the highest part often shows large values up to 2.0 (see Figure~\ref{fig-dist}(b)).
As we discuss in Sect.~\ref{ss-para-ratio}, this indicates that the Mg II lines are optically thin at upper part of spicules and the observed plasma most probably has a higher temperature ($\sim$ 20~kK) or a lower pressure ($\sim$ 0.02\pcgs).
In agreement with these findings, we show in Sect.~\ref{ss-high} that the relatively narrow Mg II profiles observed at higher altitudes, where we can expect to observe mostly individual spicules, can be reproduced by a single-slab model that does not assume a superposition of several structures along a LOS. 
As we show in the example in Figure\ref{fig-comp}(a), a suitable single-slab model can have high temperature (18~kK) and realistic micro-turbulent velocity (\Vturb\ = 10 \kms).

\par
The analysis of the observed spectra shows that at the altitudes $Y \sim$ 5\arcsec\ and below, the derived unsigned LOS velocities are small. 
The averaged values at these altitudes are lower than 5 \kms\ (see Figure~\ref{fig-dist}(c)). 
Moreover, the line profiles obtained at these altitudes are broad, with mean widths of 90 \kms\ (0.84~\AA), or higher. 
Such very broad profiles and very small unsigned LOS velocities in this part of spectra could be explained by the presence of numerous spicules with randomly distributed LOS velocities along the LOS -- hence the bulk velocity is nearly zero.
As we demonstrate in this work, multi-slab models with LOS velocities can indeed reproduce the broad and complex Mg II profiles observed at the middle heights of spicules. 
This indicates that at those heights we are likely observing a superposition of multiple dynamic spicules along any LOS.
To confirm our findings, we showed here that single-slab modeling cannot reproduce the very broad observed line profiles even with extreme parameters if we use micro-turbulent velocities below 15\kms\ (see Figure~\ref{fig-para-s}).
Large line widths can be achieved only when we introduce micro-turbulent velocities (such as 25\kms) that are higher than the local sound speed.
Even in such a case, it is difficult to reproduce the type of profiles at the middle heights that have flat-topped shape.
However, when we consider a superposition of multiple spicules along LOS that have mutually different LOS velocities, the Mg II h and k line profiles have larger widths and their tops are flat (see Figure~\ref{fig-comp}(b)).

\par
Based on the results presented in this paper, we can conclude that the width of the Mg II h and k line profiles of the off-limb spicules is strongly affected by a number of spicules present along the LOS and by their LOS motions.
We have shown that, if there is more than a single spicule along the LOS, and if these spicules have different LOS velocities, the line width increases significantly, compared to the case of a single spicule, or even multiple mutually static spicules.
The line widths of the synthetic Mg II line profiles produced by our multi-slab model with LOS velocities are up to 100\kms\ (0.93~\AA), which is quantitatively consistent with the observed Mg II line profiles at the middle ($Y \sim$ 5\arcsec) or lower altitudes ($Y \sim$ 2\arcsec).
This demonstrates that the Mg II h and k line profiles are strongly affected by the superposition effect with several spicules along the LOS, where each spicule has different LOS velocity.
Therefore, it is not adequate to assume that the observed Mg II h and k spectra include information only about the spicule at the front, even though the Mg II h and k lines are optically thick at the line center.
However, we also show that the spectra obtained at higher altitudes are less contaminated by the presence of numerous spicules along the LOS and we can assume that this spectra likely carries information only about a single or a few spicules.

\par
In the present paper, we took LOS velocities randomly from a uniform distribution between $\pm$ 25\kms.
These values are based on the observations, where the LOS component of radial velocities and swaying or torsional motions have the apparent velocity of the order of $\pm$ 25\kms.
Our concept of multi-slab models with random LOS velocities is thus consistent not only with the results of the analysis of observed spectra presented here but also with previous imaging observations of spicule dynamics.

\par 
A precise determination of the plasma properties of the observed spicules is out of the scope of the present paper. 
Therefore, we did not aim here to find a perfect one-to-one fit between the observed and synthetic Mg II line profiles. 
However, examples of comparison between the observed profiles and the profiles produced by the models show a promising agreement. 
The reason for this is two-fold. 
First, we demonstrated that the width of the Mg II h and k line profiles can be significantly influenced by a superposition of several modeled spicules with realistic dynamics along a LOS. 
Second, we showed for the first time that the intensity of the Mg II h and k lines can be influenced via the Doppler brightening/dimming effect by assuming realistic radial velocities. 
We will exploit these effects in future modeling of spicules. 
We will also take into account the mutual illumination between individual spicules \citep[e.g.][]{hei89} -- an effect that is not assumed in the present work. 
In addition, we are preparing a work on statistical inference of solar spicules using 3D stochastic radiative transfer modeling.

\acknowledgments
IRIS is a NASA small explorer mission developed and operated by LMSAL with mission operations executed at NASA Ames Research center and major contributions to downlink communications funded by ESA and the Norwegian Space Centre.
SDO is part of NASA's Living With a Star Program.
This work was supported by JSPS KAKENHI Grant Number JP17J07733 (PI: A.T.).
S.G., P.H., and S.J. acknowledge support from the grant No. 19-17102S of the Czech Science Foundation (GA \v CR).
S.G., P.H., and J.\v S. acknowledge support from the grants No. 19-16890S and 19-20632S of the Czech Science Foundation (GA \v CR). 
A.T., S.G., P.H., J.\v S., and S.J. thank for the support from project RVO:67985815 of the Astronomical Institute of the Czech Academy of Sciences.
S.J. acknowledges the financial support from the Slovenian Research Agency No. P1-0188.
A.T. thanks for the hospitality of the Astronomical Institute of the Czech Academy of Sciences during her visits and for support from the grant No. 19-17102S (GA \v CR).


\end{document}